\documentclass[letterpaper, 11pt, notitlepage]{article}
\usepackage{osameet3}

\usepackage[T1]{fontenc}
\usepackage{float}
\usepackage{graphicx}
\usepackage{epstopdf}
\usepackage{amssymb}
\usepackage{amsmath}
\usepackage[urlcolor=blue,hyperindex,colorlinks,bookmarks=true,linkcolor=black,citecolor=black,]{hyperref}
\usepackage[normalem]{ulem}
\usepackage[abs]{overpic}
\usepackage{color}
\usepackage{rotating}
\usepackage{subfigure}
\usepackage{physics}
\usepackage{bm}
\usepackage{enumerate}
\usepackage[utf8]{inputenc}
\usepackage{booktabs}
\usepackage{placeins}
\usepackage{hyperref}

\usepackage{xr}
\makeatletter

\newcommand*{\addFileDependency}[1]{
\typeout{(#1)}
\@addtofilelist{#1}
\IfFileExists{#1}{}{\typeout{No file #1.}}
}\makeatother

\newcommand*{\myexternaldocument}[1]{%
\externaldocument{#1}%
\addFileDependency{#1.tex}%
\addFileDependency{#1.aux}%
}

\myexternaldocument{SI}

\usepackage[pagewise]{lineno}
\usepackage{setspace}

\usepackage[style=phys,biblabel=superscript,chaptertitle=false]{biblatex}
\usepackage[english]{babel}
\usepackage{blindtext,tikz}
\usetikzlibrary{calc}

\usepackage{siunitx}

\begin{document}

\title{Large Area Optical Frequency Detectors\\
for Single-Shot Phase Readout}
\date{\today}
\author{\small{Felix Ritzkowsky$^{1,\dagger,*}$, Matthew Yeung$^{2,\dagger}$, Engjell Bebeti$^1$, Thomas Gebert$^{3,4}$,\\ Toru Matsuyama$^3$, Matthias Budden$^{3,4}$, Roland E. Mainz$^1$, Huseyin Cankaya$^1$,\\
Karl K. Berggren$^{2}$, Giulio Maria Rossi$^1$, Phillip~D.~Keathley$^{2,*}$, and   Franz X. Kärtner$^1$}}
\address{$^1$Deutsches Elektronen Synchrotron (DESY) \& Center for Free-Electron Laser Science, Notkestr. 85, 22607 Hamburg, Germany\\
$^2$Research Laboratory of Electronics, Massachusetts Institute of Technology,\\
77 Massachusetts Avenue, Cambridge, MA 02139, USA\\
$^3$ Max Planck Institute for the Structure and Dynamics of Matter,
Luruper Chaussee 149, 22761, Hamburg, Germany\\
$^4$ WiredSense GmbH, Luruper Hauptstr. 1, 22547 Hamburg, Germany\\
$^5$Department of Physics and The Hamburg Centre for Ultrafast Imaging,
Universität Hamburg, Luruper Chaussee 149, 22761 Hamburg, Germany\\
$^\dagger$These authors contributed equally to this work.
}
\email{$^*$e-mail: felix.ritzkowsky@desy.de; pdkeat2@mit.edu}

\maketitle

\section*{Abstract}
Attosecond science has demonstrated that electrons can be controlled on the sub-cycle time scale of an optical wave, paving the way toward optical frequency electronics \cite{krausz_attosecond_2014}. Using controlled few-cycle optical waveforms with microjoules of energy, the study of sub-cycle electron emission has enabled the generation of attosecond ultraviolet pulses and the control of attosecond currents inside of solids. However, these experiments rely on high-energy laser pulses and test systems not suitable for integration in microcircuit elements.
To move towards integrated optical frequency electronics and their practical applications, a system suitable for integration into microcircuits capable of generating detectable signals with low pulse energies is needed. While current from plasmonic nanoantenna emitters can be driven at optical frequencies, low charge yields have been a significant limitation. In this work we demonstrate that large-scale electrically-connected plasmonic nanoantenna networks, when driven in concert, enable a much higher charge yield sufficient for shot-to-shot carrier-envelope phase detection, which is a hallmark of the underlying sub-cycle processes. We use a tailored sub-2-cycle mid-infrared waveform of only tens of nanojoules of energy to drive in excess of 2000 carrier-envelope-phase-sensitive electrons from interconnected plasmonic nanoantenna arrays that we detect on a single-shot basis using conventional readout electronics. 
 Our work shows that electronically integrated plasmonic nanoantennas are a viable approach to integrated optical frequency electronics. By engineering the nanoantennas to the particular use case, such as carrier-envelope phase detection, and optimizing the density and total amount, the output signals are fully controlled. This flexible approach to optical frequency electronics will further enable many interesting applications, such as petahertz-bandwidth electric field sampling or the realization of logic gates operating at optical frequencies \cite{bionta_-chip_2021,boolakee_light-field_2022}. 

\section*{Introduction}
When John A. Fleming developed the first widely usable vacuum diode based on the thermionic emission of electrons from a tungsten filament and showed for the first time rectification of electronic AC signals, he laid the foundation for modern electronics \cite{fleming_conversion_1905}. 
Around one hundred years later, in the pursuit of ever faster frequency electronics, a major advancement was made by utilizing carrier-envelope phase (CEP) controlled few-cycle pulses to rectify electric fields at hundreds of terahertz at sharp metal tips \cite{Kruger2011}. This not only demonstrated the generation of rectified, optical-frequency currents, but also demonstrated control over attosecond electron currents by controlling the optical pulse CEP. Subsequent investigations into these emission processes revealed complex attosecond-fast dynamics \cite{kim_attosecond_2023,dienstbier_tracing_2023}.

With the goal of achieving electronics operating at the frequency of optical waves, many methods were investigated to generate rectified femto- to attosecond currents directly in closed electric circuit elements.  For example, by using sub-cycle interband transitions in dielectrics \cite{schultze_controlling_2013,schiffrin_optical-field-induced_2013,sederberg_vectorized_2020,higuchi_light-field-driven_2017}, or metallic nanoantennas \cite{rybka_sub-cycle_2016,putnam_optical-field-controlled_2017}. These steps toward integrated circuits significantly reduced the experimental requirements from large and bulky vacuum equipment to low-energy ambient operation. Applications exploiting the sub-cycle nature of these currents have been demonstrated. Examples include attosecond resolution electric field measurements, CEP detection of few-cycle pulses, and petahertz logic gates \cite{schiffrin_optical-field-induced_2013,park_direct_2018,bionta_-chip_2021,zimin_petahertz-scale_2021,hui_attosecond_2022,rybka_sub-cycle_2016,putnam_optical-field-controlled_2017,kubullek_single-shot_2020,higuchi_light-field-driven_2017,kubullek_single-shot_2020,boolakee_light-field_2022,lee_model_2018}. Resonant nanoantennas have emerged as an attractive option, as they significantly reduce the energy required for field emission by optical pulses \cite{rybka_sub-cycle_2016,putnam_optical-field-controlled_2017,yang_light_2020,bionta_-chip_2021,keathley_vanishing_2019}. This reduction can reach up to three orders of magnitude, lowering the energy requirement to picojoule levels, while confining electron emission to a well-defined hotspot at the sharp tip of the nanoantenna. Additionally, by exploiting the extreme spatial confinement of nanoantennas, attosecond time-scale charge transport across nanometer-sized junctions has been achieved \cite{ludwig_sub-femtosecond_2020}.

While resonant nanoantennas offer several advantages, they also have limitations that impact their practicality. The electron yield from these nanoantennas has typically been less than one electron per shot \cite{rybka_sub-cycle_2016,putnam_optical-field-controlled_2017,bionta_-chip_2021,yang_light_2020}. As a result, thousands of individual laser shots must be integrated to achieve a statistically significant signal, which restricts the applicability to high-repetition-rate laser sources.  Ideally, enough current would be generated per laser shot for CEP-sensitive readout without the need for averaging.  
Simply increasing the peak intensity of the laser pulse cannot scale the signal level of these devices, as this would cause irreversible laser-induced damage.  

In this work, we demonstrate scalable and sensitive CEP detection through the simultaneous excitation of hundreds of interconnected off-resonant metallic nanoantennas \cite{yang_light_2020}. This approach enables coherently-driven, attosecond-fast electron emission across the entire detector area of \SI{225}{\micro\meter\squared}. Moreover, by employing a custom-developed mid-infrared (MIR) sub-2-cycle laser source \cite{ritzkowsky_passively_2023} we obtain a more than tenfold increase in charge emission per individual antenna compared to previous results, with a CEP-sensitive charge emission as high as 3.3~e per shot per antenna \cite{yang_light_2020}. Optical pulses with longer central wavelengths have a proportionally higher electron yield per individual half-cycle compared to their shorter wavelength counterparts. Additionally, the longer wavelength driver excites the nanoantenna off-resonantly, which enables the full reproduction of the incident electric field at the nanoantenna tip. This is crucial, as the number of optical cycles dramatically influences the amount of CEP-sensitive charge produced \cite{yang_light_2020}.  
Through this combination of short-pulse exciation and scaling of the emitter area, we achieve a CEP-sensitive charge yield in excess of 2300~e per shot, achieving single-shot operation at the full repetition rate of the laser system (\SI{50}{\kilo\hertz}). The energy requirements of less than \SI{100}{\nano\joule} represent a reduction of 2 orders of magnitude compared to alternative gas-phase methods, while removing the need for vacuum conditions \cite{hoff_continuous_2018}. Such devices enable compact, shot-to-shot CEP detection for various attosecond experiments that require CEP diagnostics \cite{sola_controlling_2006,sansone_isolated_2006,rossi_sub-cycle_2020}.  Our work more broadly demonstrates the viability of low-energy, chip-scale petahertz-electronics with single-shot readout.  

\begin{figure}[ht]
    \centering
    \includegraphics[width=18cm]{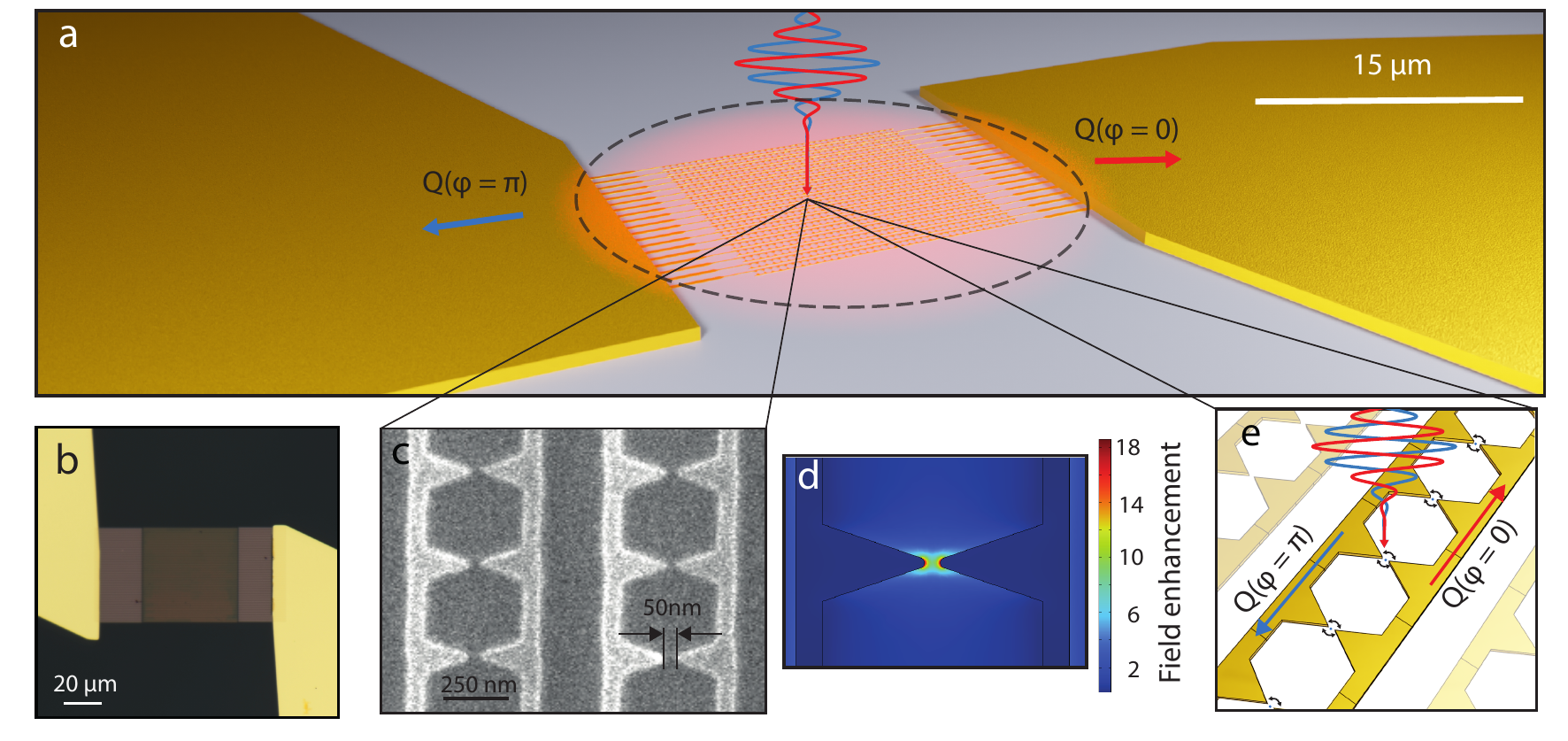}
    \caption{\textbf{CEP dependent charge generation in Nanoantenna Arrays: a}) Schematic of the charge generation process in the network showing two electric fields with a $\pi$ CEP shift corresponding to charge generated with positive $Q(\varphi = 0)$ or negative sign $Q(\varphi = \pi)$. \textbf{b}) Optical microscope image of an integrated nanoantenna network contacted with gold leads. \textbf{c}) Scanning electron microscope image of a metallic nanoantenna array. \textbf{d}) Finite-element method simulation using COMSOL of the spatial field enhancement distribution of a single antenna pair. \textbf{e}) Schematic of the nanoscopic emission process, showing the sub-cycle electron currents generated in the antenna-vacuum junction by the driving field.}
    \label{fig:Overview}
\end{figure}\section*{Designing an Optical Frequency Detector for Phase Readout}
Our devices, as seen in Fig. \ref{fig:Overview} a, consist of 722 interconnected metallic (Au) bow-tie nanoantennas embedded in a \SI{15} {\micro\meter} by \SI{15}{\micro\meter} network. The device is integrated into a macroscopic circuit, allowing for charge readout with conventional electronics. The individual bow-tie nanoantennas, as shown in the scanning electron microscope image in Fig. \ref{fig:Overview} c, have designed dimensions of \SI{530}{\nano\meter} in length, \SI{142}{\nano\meter} in width and \SI{20}{\nano\meter} in thickness, resulting in an antenna density of \SI{3.2}{\per\micro\meter\squared}. Fig. \ref{fig:Overview} d shows the finite element electromagnetic simulation of the field distribution, showing a peak enhancement of $\sim$8.2-fold for \SI{111}{\tera\hertz} (\SI{2.7}{\micro\meter} wavelength) localized at the tips of the bow-tie structure. This creates a spatially-confined hot spot for electron emission to occur. When the whole network is illuminated with a few-cycle infrared laser pulse with a peak electric field on the order of \SI{1}{\volt\per\nano\meter}, highly nonlinear tunnel ionization of electrons occurs at these hotspots at the tip of the bow-tie antennas. Additionally, the tunnel ionization is temporally confined to the peak regions of the strongest half-cycles of the exciting field. 

In the case of sufficiently strong electric fields, with a Keldysh parameter $\gamma\ll 1$ the tunneling emission for a metal-vacuum boundary is described by the quasi-static Fowler-Nordheim tunneling rate $\Gamma_{FN}(E) \propto \theta(E) \alpha \left( E \right)^2 \text{exp}\left( - \frac{f_{t}}{\vert E\vert}\right)$ \cite{keldysh_ionization_1965,bunkin_cold_1965,Fowler1928,yudin_nonadiabatic_2001}, with $\theta(E)$ noting the Heaviside function, $f_{t}=$\SI{78.7}{\volt\per\nano\meter} the characteristic tunneling field strength for gold and $\alpha$ a material and geometry dependent scaling factor. Since a single bow-tie is, in fact, a symmetric system consisting of two metal surfaces facing each other with a \SI{50}{\nano\meter} vacuum gap, we can approximate the total instantaneous currents at the junction with $\Gamma(E)=\Gamma_{FN}(E)-\Gamma_{FN}(-E)$, as experimentally shown in \cite{rybka_sub-cycle_2016,ludwig_sub-femtosecond_2020,yang_light_2020}.  A CW laser would lead to fully symmetric charge injection and transport across the gap. In this case, the time average of the residual charge in the network is zero. For the case of a few- to single-cycle pulse, however, the highly nonlinear dependence of the tunnel emission with respect to the electric field amplitude does generate a residual net charge. This is caused by the significant amplitude differences between the individual half-cycles of the pulse, effectively breaking the symmetry of emission and transport. To understand the symmetry breaking, it is useful to look at the detailed instantaneous tunneling rates as a function of the electric fields for a metal-vacuum boundary. 

\begin{figure}[ht]
    \centering
    \includegraphics[width=9cm]{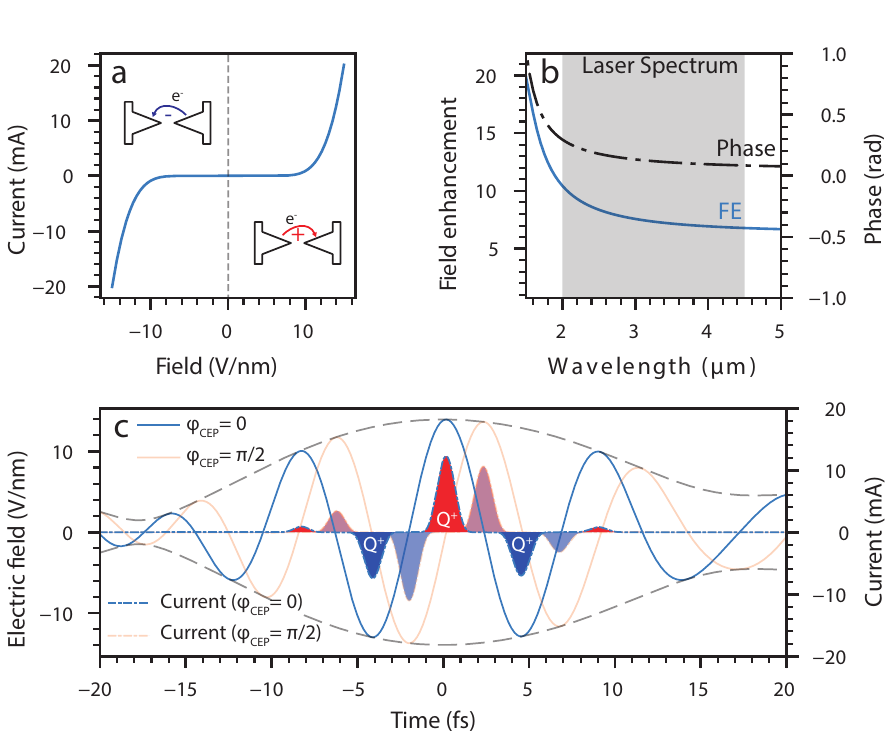}
    \caption{\textbf{Theoretical description of the antenna gap currents: a}, Effective instantaneous tunneling rate for two opposing gold surfaces in the nanoantenna junction, assuming scaling parameters from \cite{buckley_nanoantenna_2021} with an effective emission area of \SI{628}{\nano\meter\squared}. \textbf{b}, The response function of the local electric field at the tip of the nanoantenna to an exciting electric field simulated using a FEM electromagnetic solver. The simulation shows the wavelength-dependent field enhancement and phase. The effective field-enhancement of the incident pulse is $\sim 8.2$. \textbf{c}, The electric field as a function of time and the instantaneous current as a function of the electric field for a CEP of $\varphi = 0,\pi/2 $. The electric field is the calculated local antenna field using the characterized optical pulse and the simulated antenna response (see Supplementary Sec. S2 and S3.1).  The solid lines note the electric field and the dashed lines the current. The shaded areas underneath the current curves show the total charge yield, with red areas contributing positively and blue areas contributing negatively.}
    \label{fig:theory}
\end{figure}

The instantaneous response of this system, shown in Fig. \ref{fig:theory} a, is equivalent to two parallel diodes in opposing directions. Emission rates are taken from \cite{buckley_nanoantenna_2021} and use the antenna tip surface area of \SI{628}{\nano\meter\squared}. 

Regarding the electron emission process, the local field at the tip of the nanoantenna is relevant. Therefore, we need to consider the antenna's complex transfer function \cite{bionta_-chip_2021}. 
The antenna is designed to be off-resonant for two main reasons; the first is to transfer the full bandwidth of the optical pulse to the antenna tip, as a sharp resonance would increase the local pulse duration and reduce the CEP-dependent charge yield drastically. The second reason is that the fabrication process is not fully uniform throughout the detector area, resulting in small spectral shifts of the antenna resonance \cite{yang_light_2020}. When designed on-resonance, small variations will transfer to sharp phase differences between individual antennas, as the phase response has a sharp jump at resonance. Therefore, when the antennas are driven off-resonance, small variations in the fabrication will not translate into sharp phase changes of the optical field at the antenna tip.
The local field enhancement and the phase response of an off-resonance antenna for wavelengths above \SI{2}{\micro\meter} is shown in Fig. \ref{fig:theory} b. The local field at the antenna tip $E_{\text{loc.}}$ is therefore the frequency domain multiplication of the incident pulse $\Tilde{E}(\omega)$ and the antennas complex frequency response $\Tilde{H}(\omega)$, $E_{loc.}(t) = \mathcal{F}^{-1}\{ \Tilde{E}(\omega)\cdot \Tilde{H}(\omega) \}$.
The effective instantaneous current response of the system to such a pulse with a peak field of $\sim$\SI{13}{\volt\per\nano\meter} is shown in Fig. \ref{fig:theory} c. The employed optical pulse shape is the reconstructed optical pulse used in the experimental apparatus, combined with the simulated local field enhancement (see Supplementary Sec. 2 and Sec 3.1 for details). The central half-cycle with the highest field amplitudes generates the largest peak current with up to \SI{12}{\milli\ampere} for a duration of \SI{1.1}{\femto\second} (FWHM). The neighboring half-cycles generate substantially smaller currents with the opposite sign. Since conventional electronics do not support the petahertz bandwidth currents, the device acts as an integrator, and the net charge deposited by the optical pulse resides in the circuit network, similar to a photodiode. The mathematical description of these charges $Q$ as a function of the pulse CEP $\varphi$ is simply the integral over the instantaneous currents;

 \begin{align}
 Q(\varphi) = & \int_{-\infty}^\infty \Gamma(A(t)\cdot \text{cos}\left(\omega_0t +\varphi \right)) ~\mathrm{dt}\\
 Q(\varphi) =& \int_{-\infty}^\infty \Gamma_{FN}(A(t)\cdot \text{cos}\left(\omega_0t +\varphi \right)) ~\mathrm{dt}-\int_{-\infty}^\infty \Gamma_{FN}(-A(t)\cdot \text{cos}\left(\omega_0t +\varphi \right)) ~\mathrm{dt}.\\
 Q(\varphi) =& Q^+(\varphi) - Q^-(\varphi)
 \label{eq:cepyield}
 \end{align}
 
The CEP dependence of the charge now stems from the small difference of $Q^+(\varphi)$ and $Q^-(\varphi)$. For the case of a cosine pulse ($\varphi=0$) the charge yield becomes maximal, and for the case of a sine pulse ($\varphi=\pi/2$) the charge components cancel out to zero.
Based on the results shown in \cite{yang_light_2020} with 0.1~e per antenna, one can anticipate CEP-dependent charge amplitudes of around 1.4~e per antenna for the optical pulses used in our experiments and a peak field of \SI{1.7}{\volt\per\nano\meter}. The resulting charge increase is due to a reduced number of cycles (from 2.5 to 2), and the use of a longer central wavelength \cite{buckley_nanoantenna_2021}. With the known charge yield per antenna, one can extrapolate the charge yield of an array of interconnected antennas to a charge that is within the reach of reasonable detection limits.

\section*{Experiment and Results}

The optical pulses used in this work were generated with a home-built laser source based on optical parametric amplification and difference frequency generation that delivers passively CEP stable pulses with a FWHM duration down to \SI{16} {\femto \second } at a center wavelength of \SI{2.7}{\micro\meter}. The pulse energy was $>$~\SI{84}{\nano\joule} at a repetition rate of \SI{50}{\kilo\hertz} . The CEP of the laser was controlled by adjusting the pump-seed delay in the difference frequency generation stage. This was implemented by controlling the pump beam path length via a retro-reflector mounted on a piezo-actuated linear stage. For a detailed description of the source, see the methods section \ref{methods:source} and Ref. \cite{ritzkowsky_passively_2023}.

To illuminate the nanoantenna network, we focused the incident pulse down to $\sim$~\SI{21}{\micro\meter} (FWHM) with an off-axis parabola of focal length \SI{25.4}{\milli\meter}. The nanoantenna arrays were placed in the center of the focus. For shot-to-shot charge readout, we used a custom transimpedance amplifier with a gain of \SI{1}{\giga\volt\per\ampere} and a \SI{-3}{\decibel}-bandwidth of \SI{50}{\kilo\hertz} (WiredSense GmbH). The RMS noise of our detection was measured to be $\sim 1100$~e per shot. To overcome this RMS noise floor, we illuminated a network consisting of 722 antennas in a rectangular area of \SI{15}{\micro\meter} by \SI{15}{\micro\meter} to generate in excess of 1000~e per shot. 

After interaction with the nanoantenna arrays, pulse energies were measured by a pyroelectric photodetector with the same \SI{-3}{\decibel}-bandwidth of \SI{50}{\kilo\hertz} as the transimpedance amplifier. This arrangement allowed for the simultaneous recording of shot-to-shot pulse energy fluctuations. The pyroelectric detector uses an identical transimpedance amplifier to the one used for the nanoantenna read-out to ensure comparable statistics of the two signals. More details on the acquisition and digitization of the signal are given in the Supplementary Information Sec.~3.1. 

In this experiment, each dataset consists of the measured charge from the nannoantenna array and the corresponding pulse energy, recorded for around 50~000 shots (\SI{1}{\second}). In each dataset, the CEP of the laser is linearly ramped for \SI{600}{\milli\second} with a speed of $20~\pi$~\SI{}{\radian\per\second}, starting at $\sim$\SI{120}{\milli\second}. For different datasets, the pulse energy was systematically varied by more than a factor of ten.
\begin{figure}[ht]
    \centering
    \includegraphics{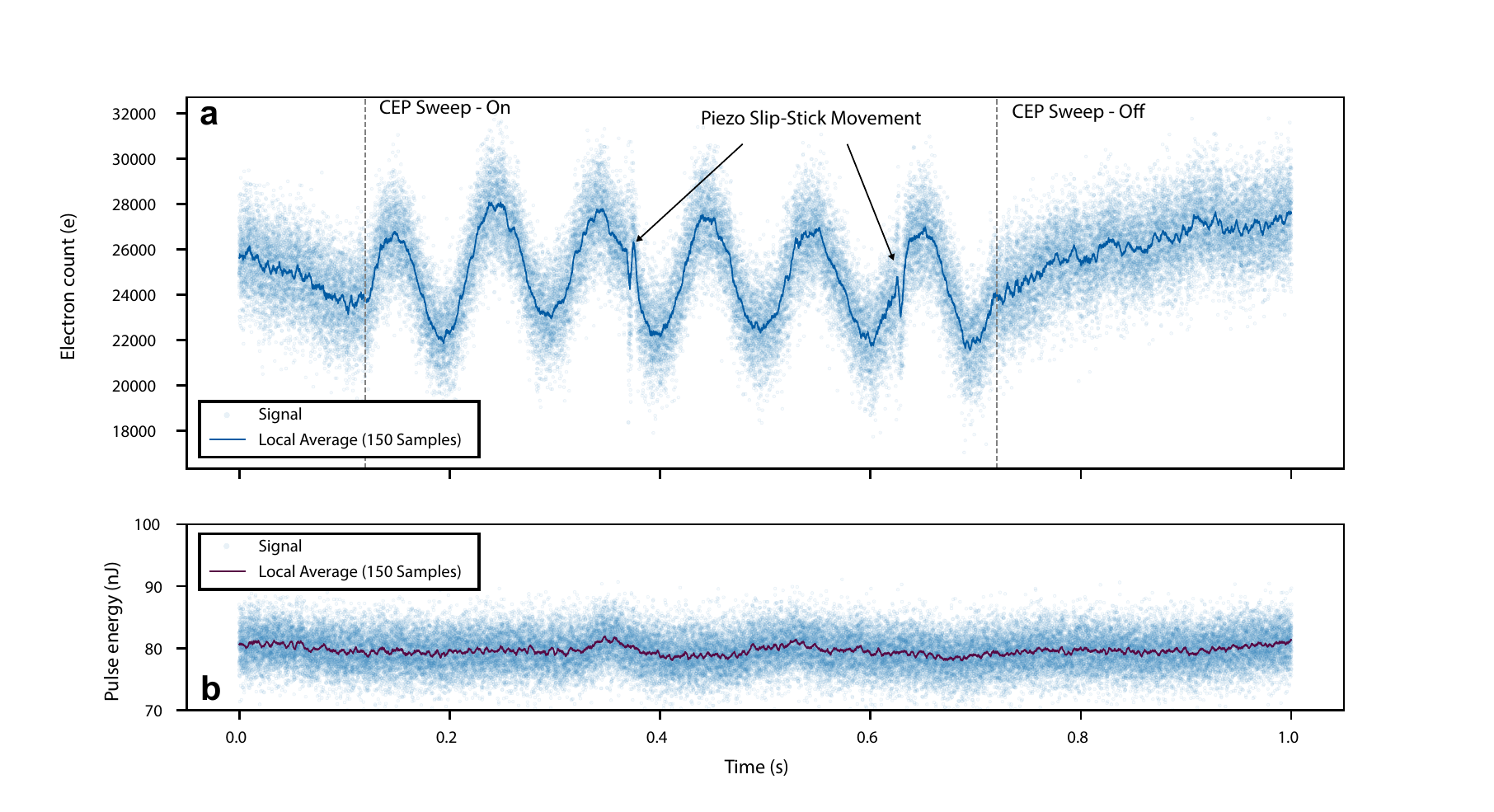}
    \caption{\textbf{Single-shot charge readout:} A single dataset recording of 50~000 laser shots for the charge yield of the nanoantenna detector (upper panel) and the laser energy recorded by the pyroelectric detector (bottom panel). The peak field of the incident laser pulse on the array is \SI{1.6}{\volt\nano\meter}. From \SI{120}{\milli\second} to \SI{720}{\milli\second} the CEP was linearly ramped over 6 cycles.}
    \label{fig:singleshot}
\end{figure}

\begin{figure}[ht]
    \centering
    \includegraphics[width=9cm]{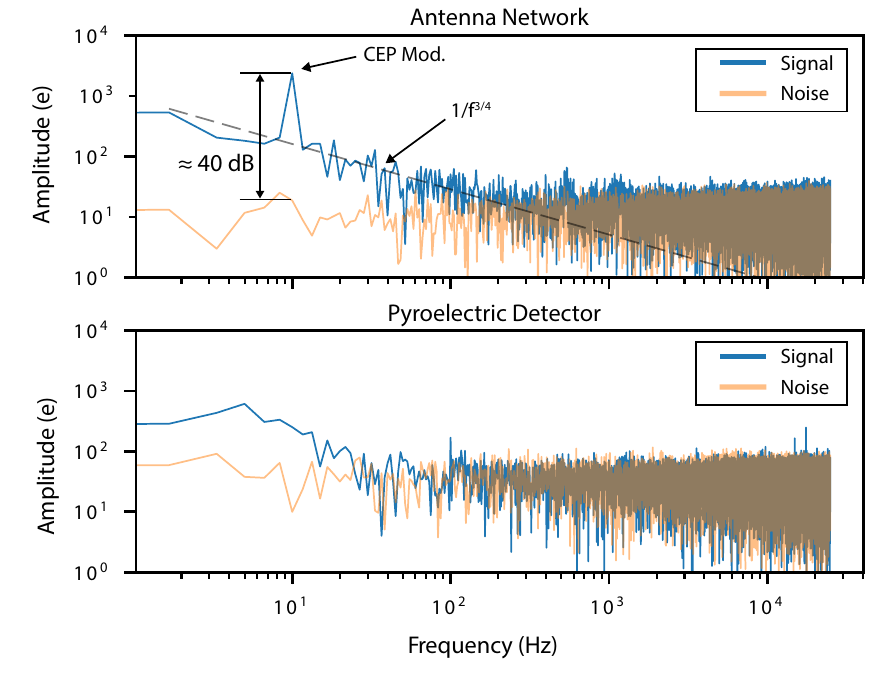}
    \caption{\textbf{Frequency Domain of the single-shot data:} The respective data from Fig. \ref{fig:singleshot} $t=~$\SI{370}{\milli\second} to $t=$~\SI{620}{\milli\second} is Fourier transformed and shown in charge amplitude as a function of frequency. For comparison, the electronic noise floor is shown in orange for both spectra. \textbf{a}) the frequency-resolved signal of the nanoantenna network. The \SI{10}{\hertz} CEP modulation is separated by \SI{40}{\decibel} from the noise floor. \textbf{b}) the frequency-resolved pulse energy fluctuation, detected with the pyroelectric detector.}
\label{fig:frequencydomain}
\end{figure}

A single dataset is presented in Fig. \ref{fig:singleshot}, including both the single-shot data and the moving average calculated over 150 shots (dark line). The upper panel shows the recorded charge produced by the nanoantenna array, with an average yield of 25~000 e per shot. From \SI{120}{\milli\second} to \SI{720}{\milli\second} the CEP is linearly ramped over a $12 \pi$ range. The data points show a clear sinusoidal CEP dependence with an amplitude of 2370~e and an signal-to-noise ratio (SNR) of 4.6, while the pulse energy does not show modulation. 
When considering the number of illuminated antennas, the individual CEP-sensitive yield per antenna and shot is 3.3~e, indicating peak currents through the nanoantenna gap of up to a 95~e/fs, corresponding to $\sim$\SI{15}{\milli\ampere}. Given the small cross-section of the nanoantenna tips, that is around \SI{628}{\nano\meter\squared}, the current density reaches a remarkable \SI{2.4}{\giga\ampere\per\square\centi\meter}.
At $t=$~\SI{370}{\milli\second} and \SI{620}{\milli\second}, sharp changes are visible in the charge yield of the detector element. These features, which are \SI{250}{\milli\second} apart, are caused by the specific movement pattern of the closed-loop slip-stick piezo stage used to control the CEP, that recenters the piezo position every \SI{1.3}{\micro\meter}.\\
To isolate the CEP-dependent signal from readout noise and pulse energy fluctuations, we Fourier transformed the dataset between $t=$\SI{120}{\milli\second} and $t=$ \SI{620}{\milli\second} and compared it to the frequency spectrum obtained without any optical input; see Fig. \ref{fig:frequencydomain}. 
The spectrum of the antenna array shows a clear peak at \SI{10}{\hertz} corresponding to the $2\pi\cdot$\SI{10}{\hertz} modulation of the CEP. This signal is around two orders of magnitude higher than the readout noise floor (corresponding to \SI{40}{\decibel}). The noise in the measured spectrum is dominated from DC to $\sim$~\SI{250}{\hertz} by $f^{-3/4}$ scaling, which is typical for field emission devices and is attributed to Brownian noise of the work function due to dynamical changes of adsorbates on the surface \cite{timm_noise_1966,yang_light_2020}. 
At frequencies higher than \SI{250}{\hertz} the spectrum is limited by shot noise, with a substantial component originating from the detection noise of the transimpedance amplifier itself. The calculated shot noise of the signal is $\sim 160$~e rms. When evaluating the recorded pulse energy fluctuations at the photodetector, no \SI{10}{\hertz} modulation is distinguishable from the background (see Fig. \ref{fig:frequencydomain} b). Above \SI{100}{\hertz} the pulse energy spectrum is dominated by detector noise. Systematic investigation of signal strength as a function of peak electric field has shown that at $\sim$~\SI{1}{\hertz} resolution bandwidth a signal distinguishable from noise can be observed down to \SI{0.6}{\volt\per\nano\meter} (corresponding to $\sim$~\SI{10}{\nano\joule}). See Supplementary Sec. 4.1 for details.

\section*{Conclusion}
We have demonstrated single-shot readout of CEP-dependent charge signals at \SI{50}{\kilo\hertz} repetition rate, underlying sub-cycle current generation across a macroscopic device area of \SI{225}{\micro\meter\squared} integrating more than $700$ individual antenna pairs. This was made possible by improving the average CEP-dependent charge yield per single antenna by a factor of $\sim$~30 \cite{yang_light_2020,ludwig_sub-femtosecond_2020}, now reaching 3.3~e per shot, and by illuminating hundreds of antennas simultaneously. The enhanced antenna yield implies a remarkable peak current density of up to \SI{2.4}{\giga\ampere\per\square\centi\meter} \cite{yang_light_2020,rybka_sub-cycle_2016,ludwig_sub-femtosecond_2020}.
With this result, we show that metallic nanoantenna networks, fabricated via state-of-the-art lithographic methods, are a flexible and scalable approach to optical frequency electronics that allows the designing of individual circuit elements, similar to conventional microelectronics. Thanks to this advance, we demonstrated off-resonant antennas that are sensitive to pulse energies two orders of magnitude lower than any other comparable single-shot absolute CEP detection technique \cite{wittmann_single-shot_2009,hoff_continuous_2018,kubullek_single-shot_2020}, enabling CEP detection of optical pulses with only tens of nanojoules of energy. Further optimization of the network density (see supplementary Sec. S2) combined with a reduced number of optical cycles in the pulse would potentially increase the total yield by an additional two orders of magnitude \cite{buckley_nanoantenna_2021,yang_light_2020}. As the measurement is dominated by read-out noise, further noise reduction of electronics downstream of the detector element will have a significant impact on SNR with potential for another 5-fold improvement \cite{andra_towards_2019}.

Given the exceptional current densities generated in these nanometer-sized devices, further studies will be necessary to elucidate the role of electron-electron interaction during the sub-cycle emission process \cite{schotz_onset_2021}.
Based on this platform, many different experiments and applications can be developed, such as the investigation of petahertz bandwidth logic gates and memory cells \cite{boolakee_light-field_2022,lee_model_2018}. The extremely small device size, comparable to the pixel size in modern Si-based CMOS detectors, combined with the reduced pulse energy requirements, enables the integration of multiple nanoantenna arrays in a larger pixel matrix. This will allow for a CEP-sensitive camera with further improved noise performance \cite{ma_photon-number-resolving_2017}. Absolute single-shot CEP tagging can also be implemented by adapting I/Q demodulation with two separate networks recording $\pi/2$ phase-shifted currents. The previously demonstrated techniques of attosecond-resolved field sampling can be extended to single-shot readout, by making large line arrays of individual networks \cite{bionta_-chip_2021,liu_single-shot_2022}. Another area of progress will be the adaptation of the fabrication process to become fully CMOS-compatible by replacing gold with aluminum or copper. With our results and natural avenues of progress, we believe that our platform will play a major role in future electronics operating at optical frequencies.

\appendix\section{Methods}

\appendix\subsection*{Laser source}\label{methods:source}
The two-cycle MIR source used to illuminate the nanoantenna networks is a home-built system based on adiabatic difference frequency generation (DFG) \cite{krogen_generation_2017} and details can be found in \cite{ritzkowsky_passively_2023}. The setup is based on a commercial Yb:KYW regenerative amplifier with a center wavelength of \SI{1.03}{\micro\meter}, a pulse duration of \SI{425}{\femto\second}, delivering up to \SI{120}{\micro\joule} at a repetition rate of \SI{50} {\kilo\hertz}. The first stage of the optical setup consists of a non-collinear optical parametric amplifier seeded with white light and pumped by the second harmonic of the pump laser \cite{manzoni_design_2016}. The amplified seed has an energy of approximately \SI{1.8}{\micro\joule} at a center wavelength of \SI{740}{\nano\meter}. After the amplification, the seed is stretched for pre-compensation of the later acquired MIR dispersion, and the pulse energy is controlled by an anti-reflection coated metallic neutral density filter wheel. For generation of the passively CEP stable DFG output in the MIR, the amplified seed and the stretched pump laser of $\sim\SI{10}{\pico\second}$ propagate collinear through an adiabatically poled Mg:LiNbO$_3$ crystal with an identical design to Krogen et al. \cite{krogen_generation_2017}. The generated broadband MIR pulse covers the spectral range of \SI{2} {\micro\meter} to \SI{4.5}{\micro\meter} at an energy of up to \SI{84}{\nano\joule} and is compressed through dispersion in BaF$_2$ and silicon. The generated pulse has a duration down to \SI{16}{\femto\second} (FWHM) at a center wavelength of \SI{2.7}{\micro\meter}, characterized by a two-dimensional spectral shearing interferometry setup. The passive CEP stability of the MIR pulse inherent to the difference frequency generation process is measured with an f-2f interferometer to \SI{190}{\milli\radian} rms over \SI{15}{\minute}.

\appendix\subsection*{Nanofabrication}\label{methods:fab}
A fused silica wafer was purchased from MTI Corporation and cut with a die saw. The substrates were cleaned by sonicating in acetone and isopropyl alcohol for 5 minutes each. Subsequently, the pieces were cleaned using an oxygen plasma. Poly(methyl methacrylate) A2 was spun at 2,500 revolutions per minute and baked at \SI{180}{\celsius}, then DisCharge H2O (DisChem Inc.) was spun at 1,000 revolutions per minute so that charging did not occur during the electron beam lithography write. 

Electron beam lithography was performed using an electron beam energy of 125 keV with doses varied from 4000-6000~\SI{}{\micro\coulomb\per\centi\meter\squared} $\mu$C cm2 with a proximity effect correction. After exposure, the resist was developed in a 3:1 isopropyl alcohol/methyl isobutyl ketone solution for 50 seconds at \SI{0}{\celsius}. Subsequently, the antenna deposition was performed using an electron beam evaporator operating below $9\cdot10^{-7}$ Torr. First, a 2 nm adhesion layer was deposited, then 20 nm of gold. Lift-off was performed in a \SI{65}{\celsius}-\SI{70}{\celsius} bath of N-methylpyrrolidone.

After antenna fabrication, contacts were patterned by photolithography using a bilayer of PMGI and S1838 both spun at 4,500 revolutions per minute. The deposition was performed by electron beam evaporation with a \SI{40}{\nano\meter} adhesion layer and \SI{160}{\nano\meter} of gold so that they could be wire-bonded to a printed circuit board. 
\appendix\section{Extended Data Figures}

\begin{figure}[ht]
    \centering  \includegraphics{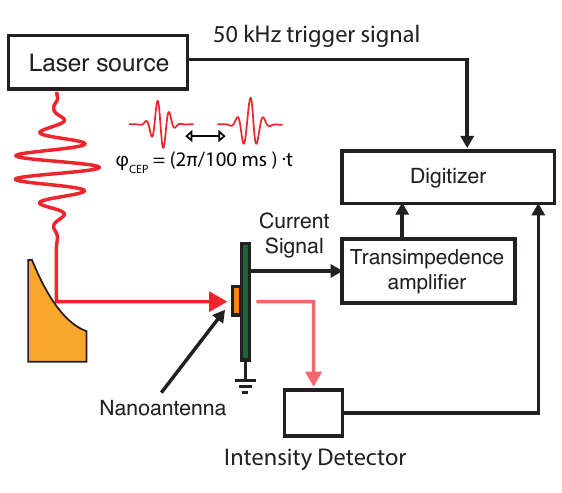}
    \caption{\textbf{Experimental setup:} The experimental setup consists out of a home-built laser source, delivering \SI{18}{\femto\second} pulses at a center wavelength of \SI{2.7}{\micro\meter} with up to \SI{84}{\nano\joule} of energy at a repetition-rate of \SI{50}{\kilo\hertz}, a \SI{25.4}{\milli\meter} focal length off-axis parabola and the nanoantenna detector element at the focal spot. For detection of the charge signal, we use a custom transimpedance amplifier with a gain of \SI{1}{\giga\volt\per\ampere} and a \SI{-3}{\decibel}-bandwidth of \SI{50}{\kilo\hertz}. For detection of the single shot intensity signal of the laser pulse, we use a \SI{50}{\kilo\hertz} bandwidth pyroelectric detector in transmission after the detector. The charge and the intensity signals are digitized with an 8-bit oscilloscope at a sampling rate of 20~MSa/s. To retrieve the individual single-shot events, the digitized pulses are integrated and sorted based on the timing signal of the \SI{50}{\kilo\hertz} trigger signal provided by the laser source. To produce a CEP-dependent signal, the CEP of the laser source is linearly swept at a rate of $2\pi / 100~\mathrm{ms}$ for \SI{600}{\milli\second}. }
    \label{fig:setup}
\end{figure}

\section*{Data Availability}
All data is available under reasonable request to the authors.

\section*{Code Availability}
All code is available under reasonable request to the authors.

\section*{Acknowledgements}
We thank John Simonaitis and Stewart Koppell for their assistance in reviewing the manuscript. The authors F. Ritzkowsky and P.D. Keathley thank William Putnam for many insightful scientific discussions on the experiments. H. Cankaya and F. Ritzkowsky thank Haim Suchowski for designing the ADFG crystal used for this work. F. Ritzkowsky and M. Yeung thank Marco Turchetti for his contributions to the sample fabrication process. M. Budden and T. Gebert thank Johannes Kunsch and Laser Components Germany GmbH  for the support in the development of pyroelectric sensor elements specifically tailored to this high-speed application.
F. Ritzkowsky and all authors affiliated with Deutsches Elektronen-Synchrotron (DESY) thank the administrative and engineering support staff at DESY for assisting with this research. We acknowledge the use of a PIER Hamburg grant for supporting M. Yeung to travel to DESY to set up initial measurements. 

F.X. Kärtner and F. Ritzkowsky acknowledge funding by: European Union’s Seventh Framework programme
145 (FP7/2007-2013) ERC Synergy Grant ‘Frontiers in Attosecond X-ray Science: Imaging and Spectroscopy’ (AXSIS) (609920); Cluster of Excellence ‘CUI: Advanced Imaging of Matter’ of the Deutsche Forschungsgemeinschaft (DFG)—EXC 2056—project ID 390715994; Deutsche Forschungs Gemeinschaft (DFG)— project ID KA908/12-1 and project ID 453615464.

This research is supported by the Air Force Office of Scientific Research (AFOSR) grant under contract NO. FA9550-18-1-0436. M. Yeung acknowledges support from the National Science Foundation Graduate Research Fellowship Program, Grant No. 1745302

\section*{Author Contributions}
FR, MY, FK and PK conceived the experiments. The samples were fabricated by MY with guidance from PK and KB. The experimental setup was constructed by FR with assistance from EB, GR, RM, and HC. The readout electronics were set up by FR, TG, MB and TM. The data was taken by FR. The data was analyzed by FR with input from MY and PK. The electromagnetic simulations were done by EB with input from FR, MY and PK. The manuscript was written by FR with significant contributions from MY, PK, GR and editing from all authors.

\section*{Competing Interests}
The authors declare no conflict of interest.
\printbibliography

\end{document}


\title{Supplementary Information for:\\
Large Scale Optical Frequency Electronics}
\author{\small{Felix Ritzkowsky$^{1,\dagger,*}$, Matthew Yeung$^{2,\dagger}$, Engjell Bebeti$^1$, Thomas Gebert$^{3,4}$,\\ Toru Matsuyama$^3$, Matthias Budden$^4$, Roland E. Mainz$^1$, Huseyin Cankaya$^1$,\\
Karl K. Berggren$^{2,*}$, Giulio Maria Rossi$^1$, Phillip~D.~Keathley$^{2,*}$, and   Franz X. Kärtner$^1$}}
\address{$^1$Deutsches Elektronen Synchrotron (DESY) \& Center for Free-Electron Laser Science, Notkestr. 85, 22607 Hamburg, Germany\\
$^2$Research Laboratory of Electronics, Massachusetts Institute of Technology,\\
77 Massachusetts Avenue, Cambridge, MA 02139, USA\\
$^3$ Max Planck Institute for the Structure and Dynamics of Matter,
Luruper Chaussee 149, 22761, Hamburg, Germany\\
$^4$ WiredSense GmbH, Luruper Hauptstr. 1, 22547 Hamburg, Germany\\
$^5$Department of Physics and The Hamburg Centre for Ultrafast Imaging,
Universität Hamburg, Luruper Chaussee 149, 22761 Hamburg, Germany\\
$^\dagger$These authors contributed equally to this work.
}
\email{$^*$e-mail: felix.ritzkowsky@desy.de; pdkeat2@mit.edu}

\maketitle
\clearpage

\section{Description of the Sub-Cycle Field Emission Current}

\begin{figure}[h!]
    \centering
    \includegraphics[width=18.4cm]{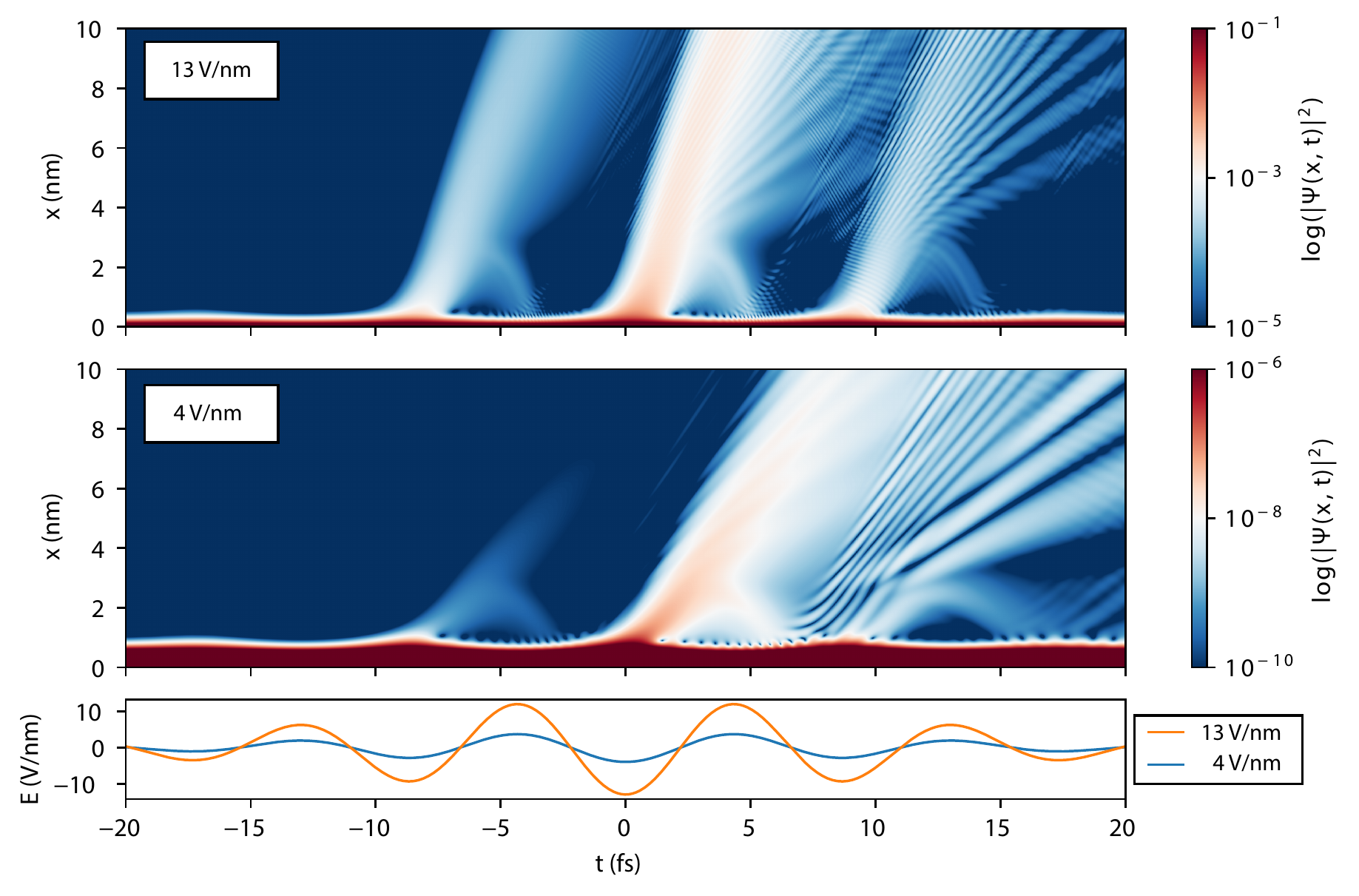}
    \caption{\textbf{Numerical Solution to the Time-Dependent Schrödinger Equation:} Integration of the time-dependent Schrödinger equation, by using the modified Crank-Nicolson scheme as described by \cite{yalunin_strong-field_2011}. The top two panels show the probability amplitudes for two cases assuming an \SI{18}{\femto\second} duration MIR pulse at a center wavelength of \SI{2.7}{\micro\meter} with \SI{4}{\volt\per\nano\meter} and \SI{13}{\volt\per\nano\meter} field strength. The bottom panel is showing the used electric field as a function of time.
    \label{figS:tdse}}
\end{figure}

To gain in qualitative understanding in the sub-cycle dynamics of the field emission process, it is useful to simulate the probability of measuring an electron at a given point in time and space outside the metal in a vacuum state. To that end, Yalunin et al. showed that the numerical integration of the time-dependent Schrödinger equation \cite{yalunin_strong-field_2011},
\begin{equation}
    i\frac{\partial \Psi}{\partial t} = \left( -\frac{1}{2}\frac{\partial^2}{\partial x^2}+V\right)\Psi,
\end{equation}
describing the interaction of a bound electron in a metal with a time-periodic field $F(t)$ is a valid approach. The potential $V$ used is of the form,
\begin{equation}
V = -
\begin{cases}
      xeF(t), & x\geq 0 \\
      E_F+W, & x < 0,
\end{cases}
\end{equation}
were $E_F$ is the Fermi energy and $W$ the workfunction. By using the modified Crank-Nicolson scheme as described in \cite{yalunin_strong-field_2011}, we calculate the probability $\vert \Psi(x,t)\vert^2$ for a gold workfunction of $W=5.1$~eV and an \SI{18}{\femto\second} duration MIR pulse at a center wavelength of \SI{2.7}{\micro\meter}. The used field strength are \SI{4}{\volt\per\nano\meter} and \SI{13}{\volt\per\nano\meter} to reflect the local fields strength measured in Fig. 5 in the main text. 

The results of the integration, presented in Fig. \ref{figS:tdse}, show the probability $\vert \Psi(x,t)\vert^2$ of measuring an electron at a given coordinate $(x,t)$. As one can see, at the low field strength of \SI{4}{\volt\per\nano\meter} the emission process is already highly sub-cycle. Driven by the peak of the electric field the electrons are ejected from the surface within a half-cycle. Strong scattering of electrons re-accelerated to the potential barrier occurs between \SI{5}{\femto\second} and \SI{10}{\femto\second}. For the case of \SI{14}{\volt\per\nano\meter} we can see stronger emission probabilities with suppressed quiver motion and rescattering at the surface. Both cases show that for the full field strength range explored in the main text in Fig. 5, the emission mechanism is still sub-cycle field-emission.\\
Despite the well described qualitative and also quantitative scaling laws of electron emission using numerical integration \cite{yalunin_strong-field_2011}, we found difficulties adapting this scheme when accounting for a system that consists of two individual emitters, driven with a $\pi$ phase-shifted pulse and subtracted as described in Eq. 3. As the CEP dependent charge is on the order of 0.1 or less of the total emitted charge, errors in the charge calculation are enlarged for the difference measurement. Therefore, we use for the CEP dependent charge yield model the quasi-static Fowler-Nordheim approximation \cite{Fowler1928}.

\section{Electromagnetic Simulation of the Nanoantenna}
\begin{figure}[ht]
    \centering
    \includegraphics[width=10cm]{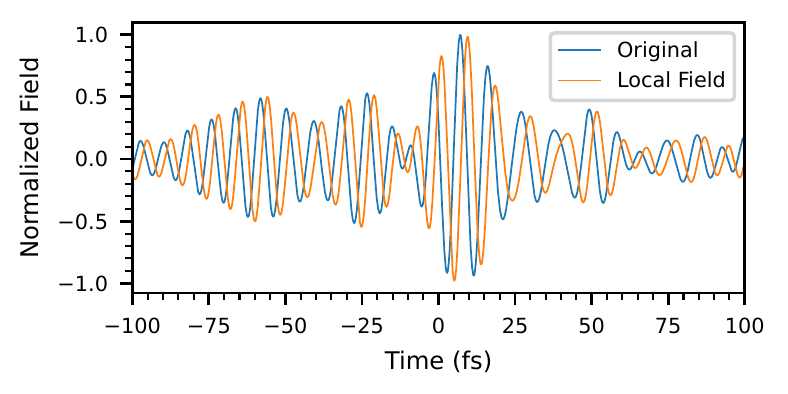}
    \caption{\textbf{Comparison of the Original Electric Field vs. the Local Field:} The original incident electric field and the calculated local field at the tip of the nanoantenna as a function of time.
    \label{figS:NLvsL}}
\end{figure}
Electromagnetic simulation was performed to estimate the local field at the apex of the nanoatenna. A simulation procedure similar to the one described in \cite{yang_light_2020} was used. A fully linear response of the device is assumed, which allows for calculation of the response function in the frequency domain. A numerical solution of the Maxwell equations is obtained using the finite element method electromagnetic waves, frequency domain solver from the wave optics module of COMSOL Multiphysics. The system was modeled by a connected antenna bow-tie consisting of gold placed on a glass substrate. The dimensions of the antenna geometry were chosen to fit the fabrication design parameters. The n and k values of gold were taken from \cite{rakic_algorithm_1995} and a constant refractive index of 1.46 was assumed for the glass substrate. An incident plane wave with a propagation direction perpendicular to the antenna-substrate interface was added on top of the geometry. The incident light is linearly polarized with the electric field being orthogonal to the connecting wires. Periodic boundary conditions were added around the antenna boundary to model the array. The semi-infinite vacuum and substrate were modeled using perfectly matched layers on top and bottom of the simulation domain. The linear response function was evaluated by comparing the results obtained with the results of an empty simulation domain with the same simulation settings.\\

\begin{figure}[ht]
	\centering
	\includegraphics[width=10cm]{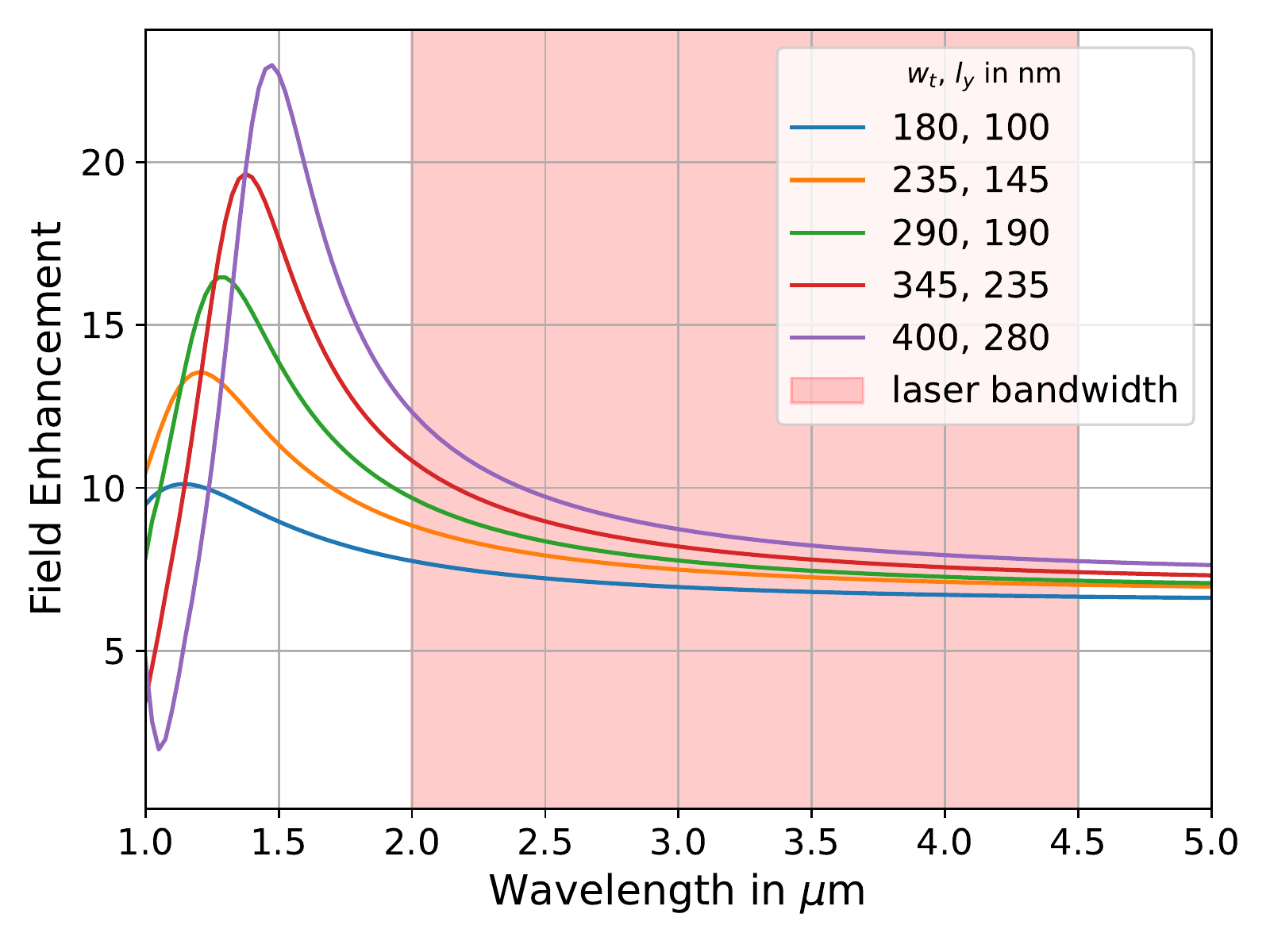}
	\caption{\textbf{Simulated Field Enhancement for Different Antenna Densities:} The average field enhancement at the nanoantenna apex as a function of wavelength for different antenna densities. $w_t$ denotes the distance between neighboring antennas and $l_y$ the closest distance between two arrays. The values $w_t = \SI{400}{\nm}$ and  $l_y = \SI{280}{\nm}$ correspond to the fabricated device presented in the main text. The most dense case with $w_t = \SI{180}{\nm}$ and  $l_y = \SI{100}{\nm}$ has an approximately $2.9$ times higher antenna density, while maintaining a similar off-resonant field enhancement.
	\label{figS:FEdensitySweep}}
\end{figure}

With the simulated complex frequency response $\Tilde{H}(\omega)$ , see Fig. 2, and the incident field $E(t)$ the local field $E_{\text{L}}(t)$ averaged over the surface of the nanoantenna tip can be calculated,
\begin{equation}
    \Tilde{E}_{\text{L}}(t) = \mathcal{F}^{-1} \{\Tilde{E}(\omega)\cdot \Tilde{H}(\omega)\}.
\end{equation}
The resulting normalized local field is shown in Fig. \ref{figS:NLvsL}. The local field is only marginally different from the incident electric field. However, the effective field-enhancement is 8.2, making the local field substantially stronger than the incident one.
Compared to the field enhancements of around 20 in  references \cite{bionta_-chip_2021,yang_light_2020,rybka_sub-cycle_2016,ludwig_sub-femtosecond_2020,putnam_optical-field-controlled_2017}, the antenna was designed to be off-resonant to preserve the incident electric field shape, while still having a sufficiently large field enhancement. Furthermore, the antenna design allows for a high antenna density of $\sim$~\SI{3}{\per\square\micro\meter}, compared to an antenna design that is resonant with the MIR field, since these would require roughly double the antenna size.\\ Further investigation of the design showed that the maximum value of the CEP dependent current is further improved by a factor of two by increasing the antenna density. The increased density damps the resonant part of the response function, but maintains a broadband off-resonant field enhancement with a factor of $\sim 6-7$. The gradual change of antenna density is shown in Fig. \ref{figS:FEdensitySweep}.  Using the formulas of the quasi static model, the increase in CEP dependent current per unit area is estimated and corresponds to a factor of two improvement from the device presented in the main text.

\newpage
\section{Experimental Setup}
\subsection{Laser Source Characterization}
\begin{figure}[ht]
    \centering
    \includegraphics[width=12cm]{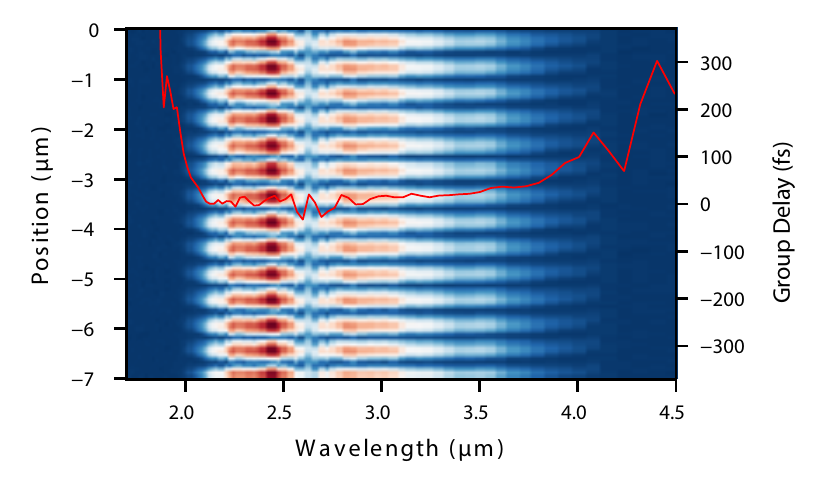}
    \caption{\textbf{Two-Dimensional Spectral Shearing Interferometry:} Measured 2DSI trace shown in logarithmic color coding. On the right hand axis in red is as an overlay the retrieved group delay.
    \label{figS:2dsi}}
\end{figure}

The laser source used to run the experiments is described in detail in reference \cite{ritzkowsky_passively_2023}. But to reflect the state of the laser source used for the described experiment, we show here the characterization of the pulse duration within a \SI{24
}{\hour} time window to the experiment. The pulse duration is measured by an adapted version of two-dimensional spectral shearing interferometry (2DSI) \cite{birge_two-dimensional_2006-1,birge_theory_2010}, where the ancillary pulses are derived from the pump laser at \SI{1.03}{\micro\meter} instead of directly from the mid-infrared pulse under test\cite{ritzkowsky_passively_2023}. This allows for the use of broadly available and cost effective spectrometers based on Si detector arrays, since in this implementation the up-converted spectrum will cover the NIR from \SI{600}{\nano\meter} to \SI{900}{\nano\meter}. The ancillary pulses are generated by fine-tuned narrowband line filters in a michelson interferometer, resulting in a shear frequency of \SI{1.35}{\tera\hertz}. The resulting measurement is shown in Fig. \ref{figS:2dsi} in logarithmic color coding. The retrieved group delay is shown on the right-hand y-axis. As can be seen the group delay is reasonably flat and shows a sharp oscillation at \SI{2.7}{\micro\meter}, which corresponds to known water absorption lines \cite{gordon_hitran2020_2022}. 

\begin{figure}[ht]
    \centering
    \includegraphics[width=10cm]{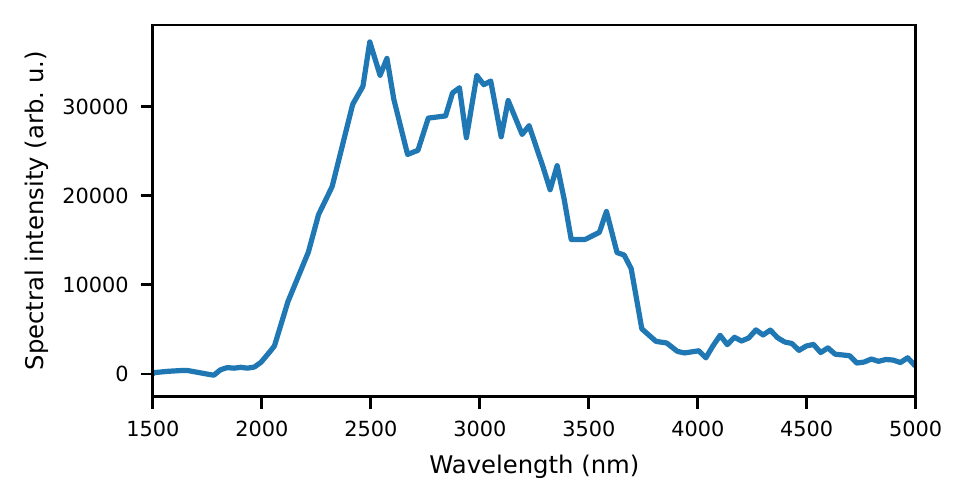}
    \caption{\textbf{Measured Mid-Infrared Spectrum:} Raw 
 mid-infrared spectrum measured on a PbSe-CCD spectrometer.
    \label{figS:spectrum}}
\end{figure}

\begin{figure}[ht]
    \centering
    \includegraphics[width=16cm]{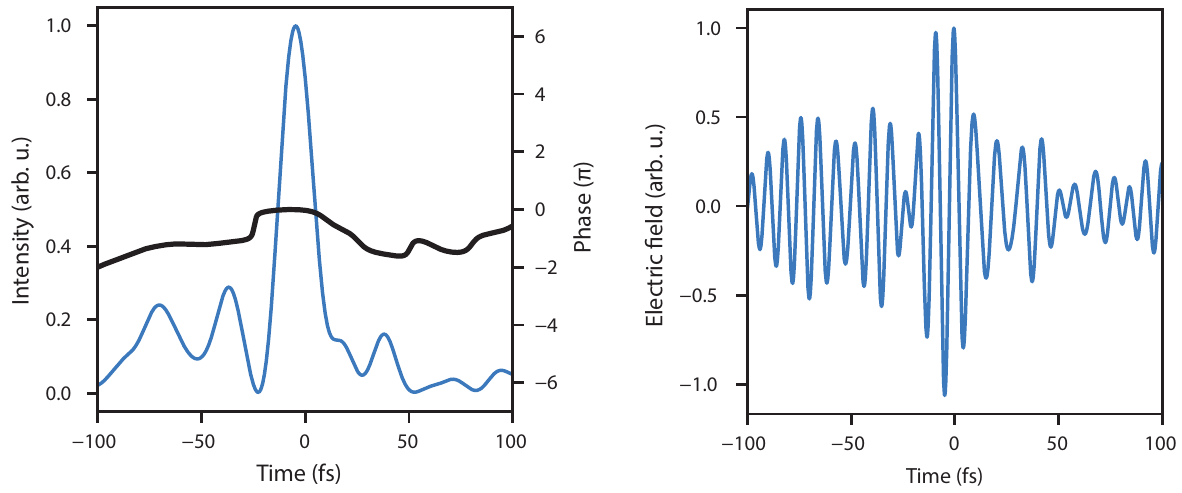}
    \caption{\textbf{Reconstructed Time Domain:} (Left) Reconstructed intensity distribution of the mid-infrared pulse (blue) and the retrieved phase (black). (Right) Electric field profile of the retrieved mid-infrared pulse set at an arbitrary CEP.)
    \label{figS:TDpulse}}
\end{figure}

In Fig. \ref{figS:spectrum} the raw mid-infrared spectrum is shown. The spectrum was measured on a PbSe-CCD spectrometer.
Using the measured mid-infrared spectrum and the retrieved group delay, the time domain of the pulse can be calculated up to an arbitrary CEP. The calculated time domain in intensity and electric field are shown in Fig. \ref{figS:TDpulse}. The retrieved pulse FWHM duration is \SI{18}{\femto\second} at a center wavelength of \SI{2.69}{\micro\meter}. This corresponds to two cycles of the carrier wave within the FWHM duration.

The passive CEP stability of the system was measured with $f$-$2f$ interferometry to be lower than \SI{190}{\radian} RMS, with details found in Ref. \cite{ritzkowsky_passively_2023}.
The CEP was controlled by adjustment of the pump-seed delay of the adiabatic difference frequency generation stage. To move the CEP by $2\pi$ of the mid-infrared pulse, the pump is delayed by one wavelength $\lambda=$\SI{1.03}{\micro\meter}. The delay is produced by a slip-stick piezo stage (SLC-2430, Smaract GmbH) and a hollow roof mirror, which delays the pump. 

\subsection{Charge Generation and Readout}\label{suppl:DAQ}

\begin{figure}[h!]
    \centering
    \includegraphics[width=12cm]{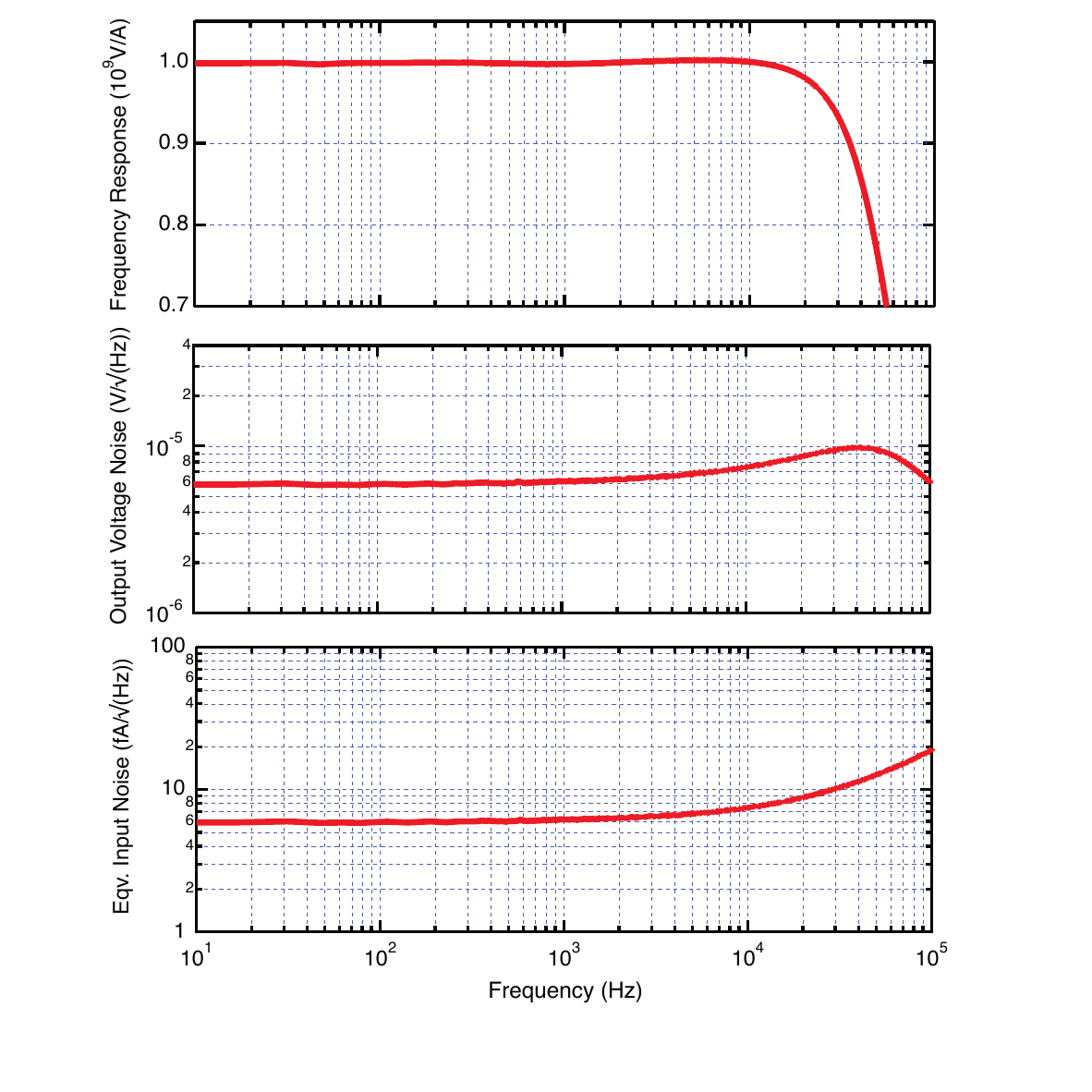}
    \caption{\textbf{Characterization of the Transimpedance Amplifier: }The TIA is characterized by measurements with an FFT analyser. (Top panel) The frequency response curve of the TIA. (Middle panel) Measured output voltage noise. (Bottom panel) Equivalent input current noise, this is calculated by dividing the output voltage noise by the frequency response of the TIA. 
    \label{figS:TIA}}
\end{figure}

\begin{figure}[h!]
    \centering
    \includegraphics[width=8cm]{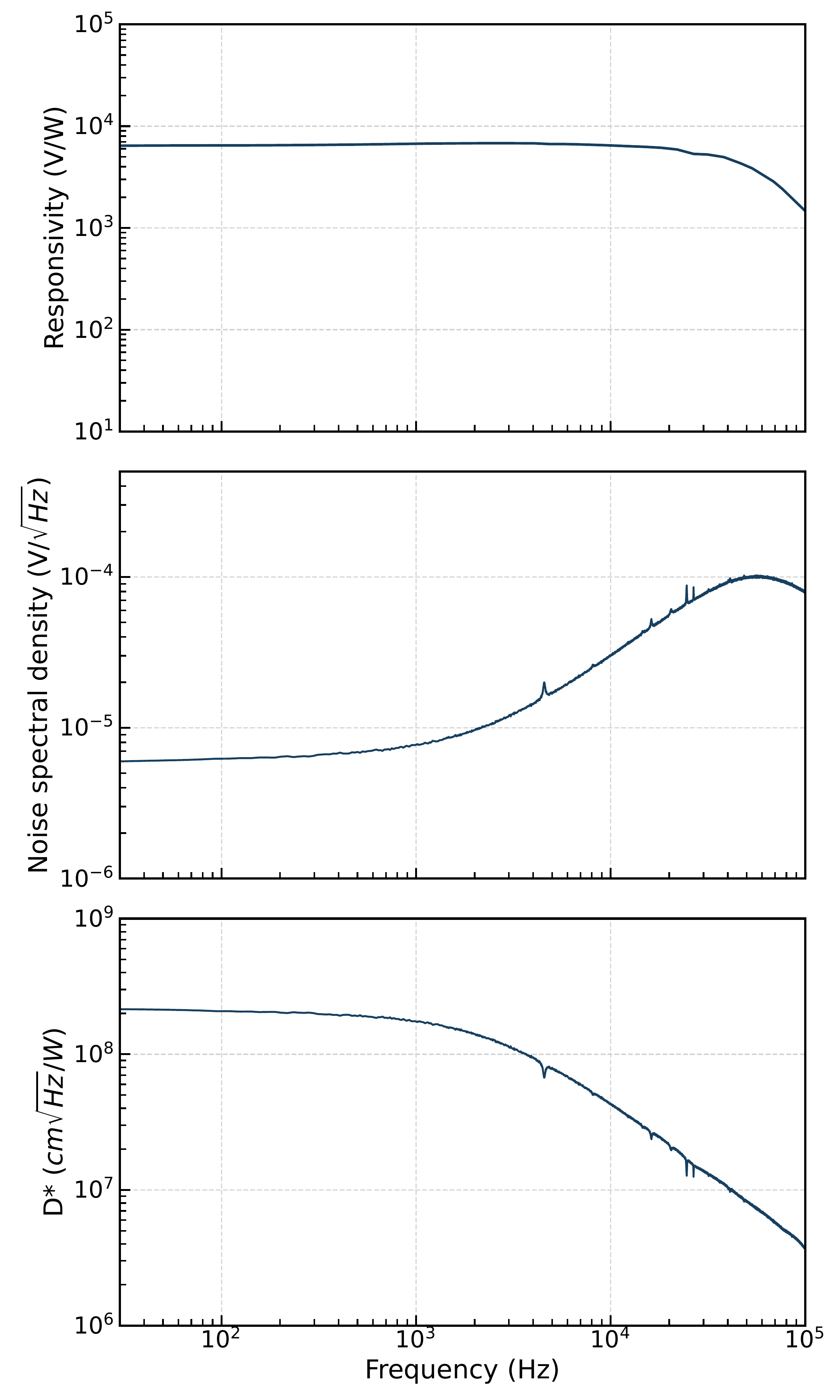}
    \caption{\textbf{Characterization of the Pyroelectric Detector:} The detector is characterized by measurement with an FFT analyser and a calibrated black body radiator at \SI{150}{\celsius}. (Top panel) The frequency response of the detector. (Middle panel) Noise spectral density of the detector with blocked input signal. (Bottom panel) Detectivity of the detector. 
    \label{figS:Pyro}}
\end{figure}

\begin{figure}[h!]
    \centering
    \includegraphics[width=10cm]{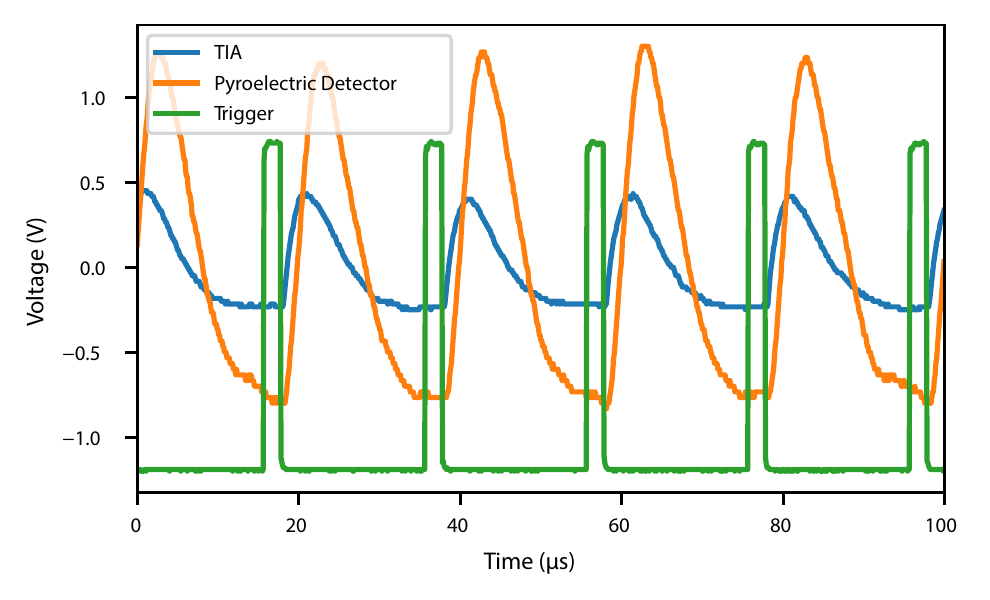}
    \caption{\textbf{Recorded Oscilloscope Trace:} Exemplary recorded \SI{100}{\micro\second} voltage time trace showing the trigger signal (green), the TIA output (blue) and the pyroelectric detector output (orange).
    \label{figS:Scope}}
\end{figure}

\begin{figure}[h!]
    \centering
    \includegraphics[width=18cm]{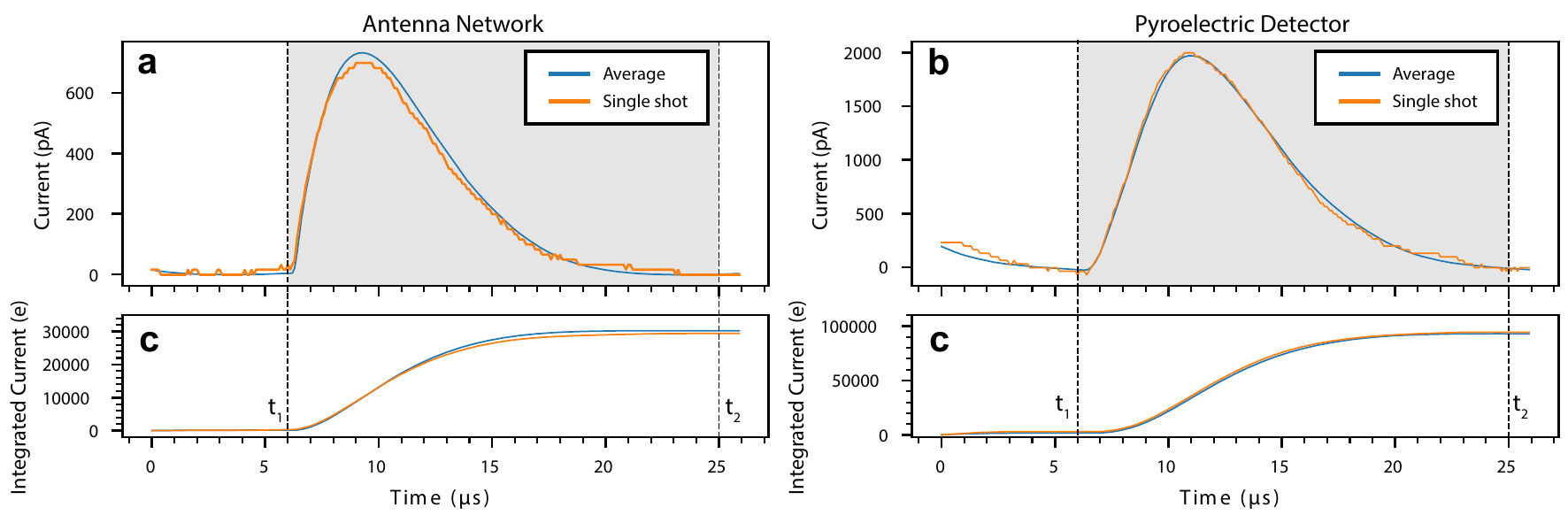}
    \caption{\textbf{Charge Integration Scheme:} (\textbf{a}) Antenna network current pulse, shown for a single pulse and averaged over many pulses to reduce noise and digitization errors. (\textbf{b}) Pyroelectric detector current pulse, showing a single pulse and the averaged pulse form. (\textbf{c}) Integrated current of the antenna network as a function of time showing a single shot and the averaged pulse form. $t_1,t_2$ denote the sampling points used for the charge calculation. (\textbf{d}) Integrated current of the pyroelectric detector pulse, showing a single shot and the averaged pulse form. $t_1,t_2$ denote the sampling points used for the charge calculation.
    \label{figS:ChargeIntegration}}
\end{figure}

For the readout of the charge in the nanoantenna network we used a custom transimpedance amplifier (TIA) provided by WiredSense GmbH. The amplifier has a total gain of \SI{1}{\giga\volt\per\ampere} at a \SI{-3}{\decibel}-bandwidth of \SI{50}{\kilo\hertz}. The input and the output of the TIA are AC coupled. The full response function, the output voltage noise and the equivalent input noise of the TIA are shown in Fig. \ref{figS:TIA}.
The pyroelectric detector used to detect the shot-to-shot intensity changes was also provided by WiredSense GmbH and it's frequency response, noise equivalent input power and detectivity are shown in Fig. \ref{figS:Pyro}. The detector was characterized with a calibrated black body radiator. To digitize both signal channels and retrieve the individual charge yields, we used an 8-bit oscilloscope recording the TIA, the pyroelectric detector and a trigger signal provided by the laser source at a sampling rate of 10~MSa/s. For the \SI{50}{\kilo\hertz} repetition rate signal, the sampling rate was chosen to provide sufficient oversampling to alleviate digitization problems of the low bit-rate oscilloscope. Oversampling by a factor of $x$ and integration of a digital signal increases the effective bit-range $n$ by $n=\ln(x)/(2\ln(2))$, if the lowest bit is submitted to sufficient Gaussian noise. An example of recorded oscilloscope trace is shown in Fig. \ref{figS:Scope}

The trigger signal falling slope was used as a reference to sort the individual shots with their respective time stamp. Before integrating over the AC coupled current signals, a baseline was introduced by averaging over the signal for \SI{3}{\micro\second} before every trigger and subtracting it locally for the respective time windows. The result of that subtraction is shown in Fig. \ref{figS:ChargeIntegration}. 
To retrieve the charge contained within each current pulse, the current pulse was integrated over and the integrated signal is sampled at points $t_1,t_2$. The integrated charge is simply the difference of charges measured  at the sampling points, $Q_{Shot}=Q(t_2)-Q(t_1)$. This technique is called correlated double sampling (CDS) and is commonly used in charged-coupled device readout circuitry\cite{oliaei_noise_2003}. The time correlated differentiation significantly reduces uncorrelated low frequency noise\cite{wey_noise_1986}.

\subsection{Device Layout}

\begin{figure}[h!]
    \centering
    \includegraphics[width=16cm]{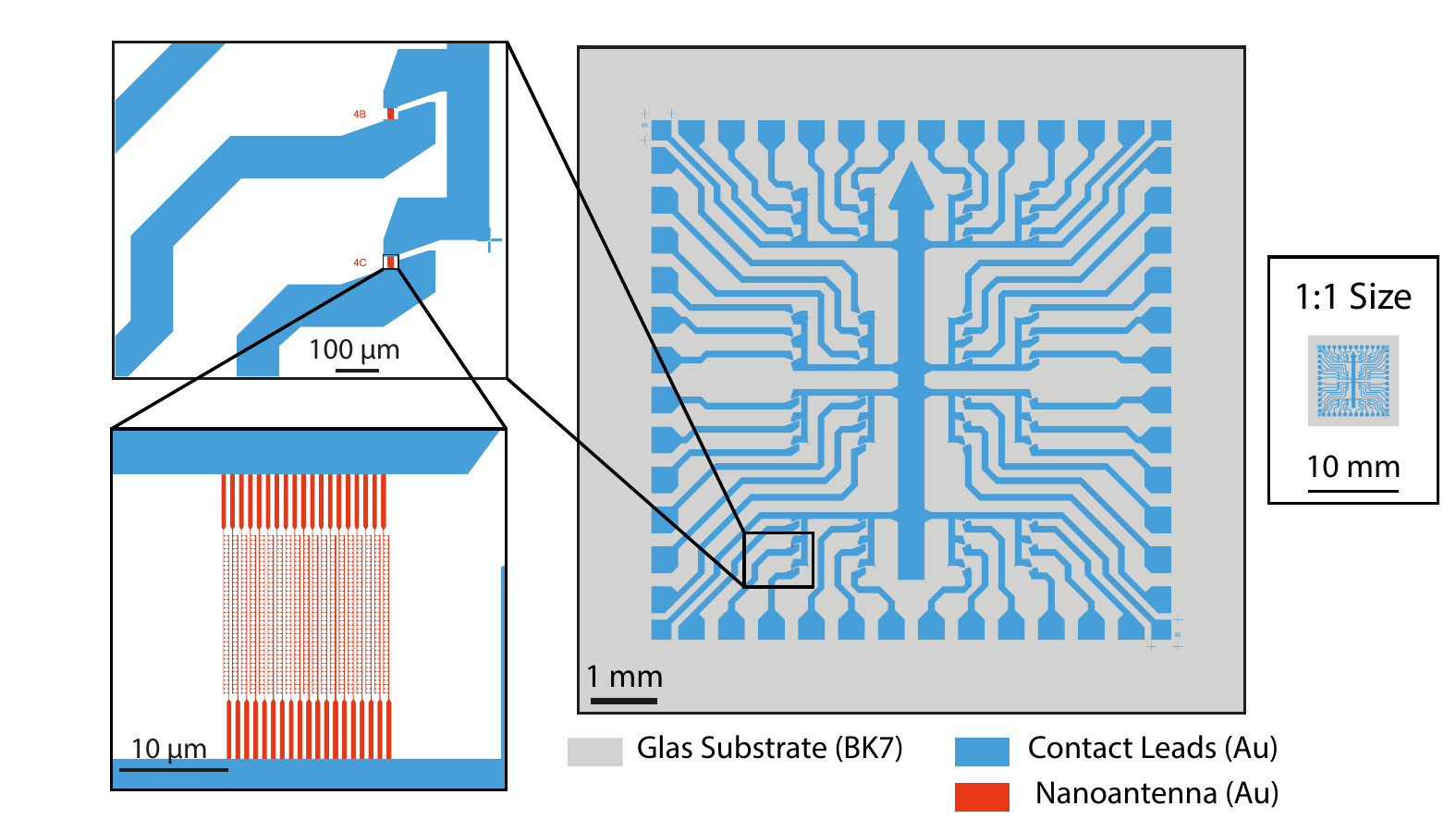}
    \caption{\textbf{Circuit Layout:} Layout of the microchip carrying the nanoantenna arrays. Progression of sizes from a 1:1 scale down to 1:1000 (Assuming Din-A4 printout of this page). Blue areas mark the contact leads fabricated through photolithography in gold (Au). Red areas mark the nanoantenna arrays fabricated through electron-beam lithography in gold (Au). Grey arrea marks the glas (BK7) substrate.
    \label{figS:layoutChip}}
\end{figure}
Fig. \ref{figS:layoutChip} shows the complete layout of the tested chip that contains the nanoantenna arrays. The chip layout is shown on a 1:1 scale with zoom-in on the relevant array tested in the main text. The device, as stated in the Methods Section \textit{Nanofabrication}, is fabricated through a two-step process. The small-scale nanoantenna arrays are fabricated on a BK7 substrate through electron-beam lithography in gold (Au). The second step is the fabrication of larger-scale contact leads through photolithography in gold (Au). The role of the contact leads is to make robust electrical contact with the nanoantenna arrays and to provide large pads at the outer edge of the chip for wire bonding.
By visual analysis in a scanning electron microscope, the best 24 arrays of 48 are selected for wire bonding to a printed circuit board.
\begin{figure}[h!]
    \centering
    \includegraphics[width=16cm]{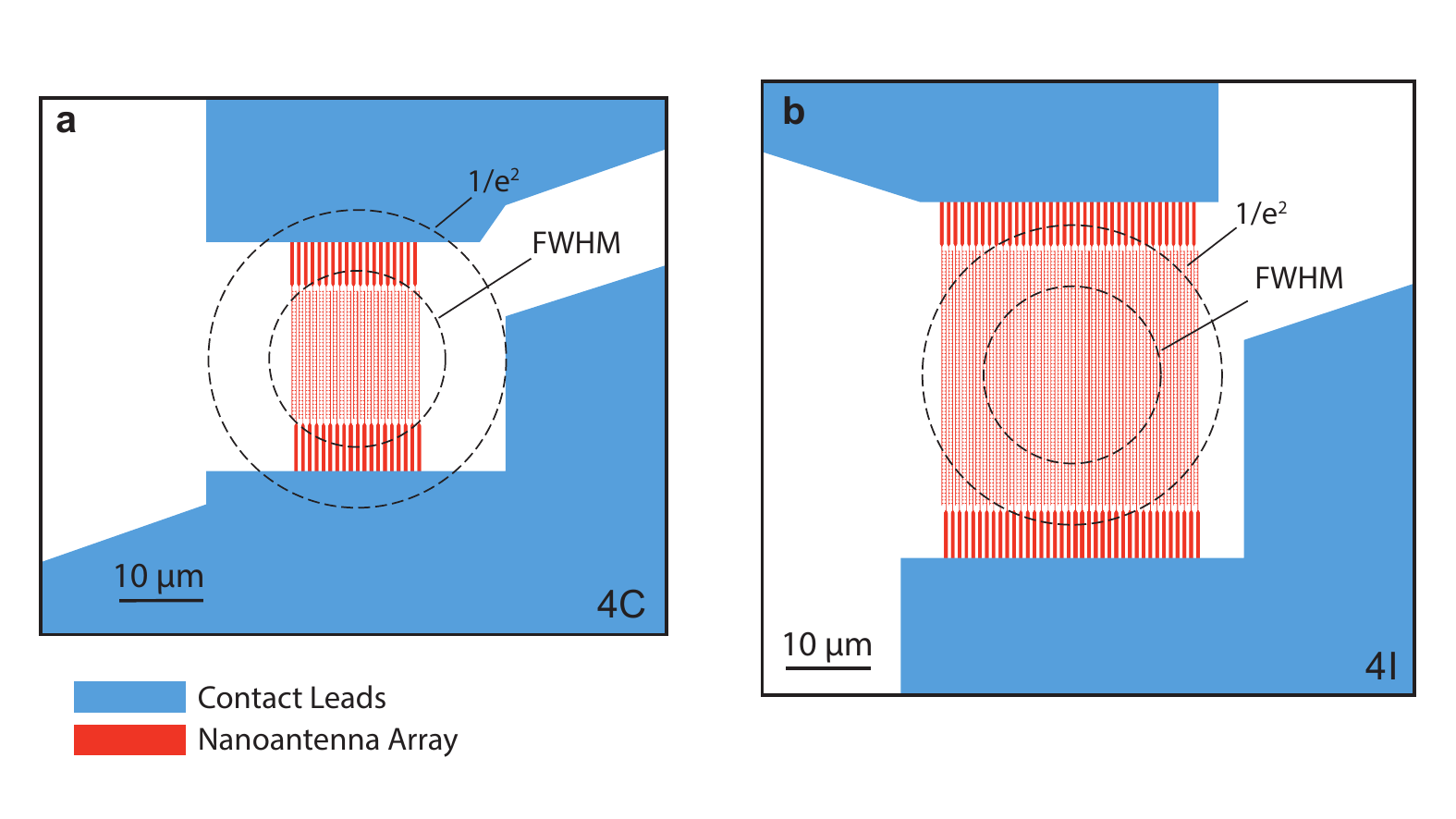}
    \caption{\textbf{Device Layout:} Layout of the devices used in this work. The blue area shows the contact leads fabricated by photolithography. The read areas show the nanoantennas made by electron-beam lithography.  The dashed circles mark the approximate spatial distribution of the laser focus (FWHM: \SI{21}{\micro\meter}, $1/e^2$: \SI{35.5}{\micro\meter}). (\textbf{a}) Layout of the device shown in the main text. The nanoantenna array area of the device measures \SI{15}{\micro\meter} by \SI{15}{\micro\meter}. (\textbf{b}) Layout of the device shown in the supplementary. The nanoantenna array area of the device measures \SI{30}{\micro\meter} by \SI{30}{\micro\meter}. 
    \label{figS:layout}}
\end{figure}
The layout of the device presented in the main text is shown in Fig. \ref{figS:layout} a. The nanoantenna array itself is shown by the red shaded structure and is produced through electron-beam lithography. The pink area is showing contact leads that connects the nanoscale devices with the large wire bond pads at the edges of the chip. The device measures \SI{15}{\micro\meter} by \SI{15}{\micro\meter} for the nanoantenna array. For comparison the spatial dimension of the optical focus (FWHM and $1/e^2$) is shown as dashed circles. An additional device with an area of \SI{30}{\micro\meter} by \SI{30}{\micro\meter} is shown in Fig. \ref{figS:layout} b. The two devices show the case of an array smaller than the laser focus and larger than the laser focus.

\newpage
\section{Complementary Measurements}
\subsection{Field Dependent Scaling of the CEP-Sensitive Charge Yield}
\begin{figure}
    \centering
    \includegraphics{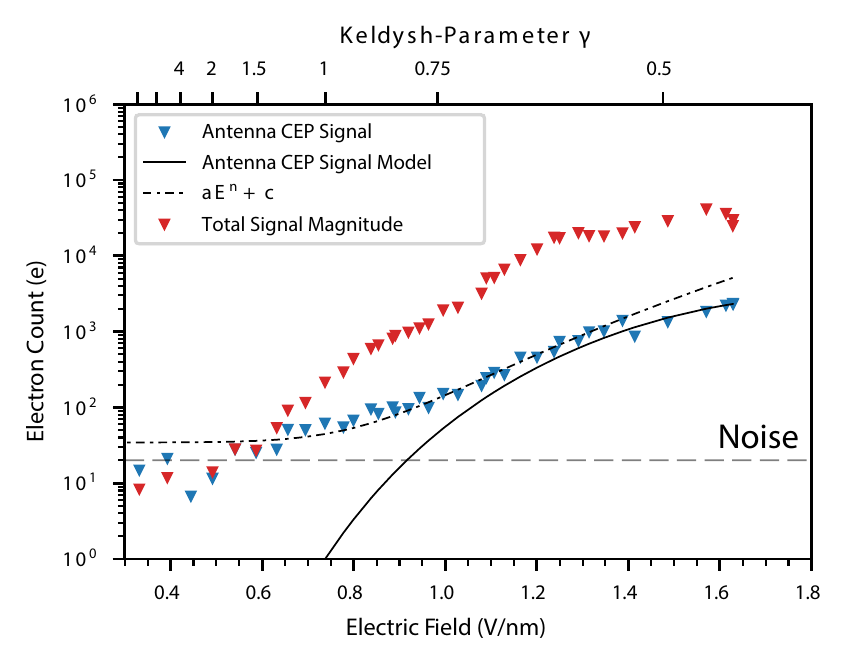}
    \caption{\textbf{Charge yield scaling:} The CEP modulation amplitude and the total magnitude (average of each measurement) of the electron counts are plotted as a function of the peak field (average of each measurement), estimated for CEP$=0$, for the respective dataset. The CEP signal is taken from the amplitude of the \SI{10}{\hertz} frequency component of the measurement data. The antenna CEP signal model uses the model described by Eq. 3 in the main text. Furthermore, a power law fit, $aE^n+c$ to the first 30 values is shown.  }
    \label{fig:scaling}
    
\end{figure}

To elucidate the mechanism underlying the CEP-dependent electron emission, the amplitude of the CEP modulation is shown in Fig. \ref{fig:scaling} (blue triangles) as a function of the incident peak field of the laser pulse.
In order to estimate the CEP-dependent emission, we first calculated the field at the tip by convolving the applied field (as retrieved by optical pulse characterization) with the calculated impulse response function of the nanoantenna. For peak fields larger than \SI{1.2}{\volt\per\nano\meter}(estimated for CEP$=0$), corresponding to a Keldysh parameter $\gamma \sim 0.6$, the CEP-dependent charge yield scales according to the quasi-static tunneling approximation shown in Eq. 3 (main text).
For values below \SI{1.2}{\volt\per\nano\meter}, the data follows a power law model $aE^n + c$, with $n=7.85$, $a=110$ and $c=35$. This scaling behavior suggests a transition from nonadiabatic tunneling emission to the quasi-static tunneling regime \cite{yudin_nonadiabatic_2001,shi_femtosecond_2021}. This scaling behavior was verified by repeating the experiment with a different nanoantenna array. The results of this second device are presented here in Supplementary Sec. S4.2. The interaction of the optical pulse with our nanoanntenna arrays generates not only CEP-dependent charges but also a pulse energy-dependent charge offset. The magnitude of the average charge yield of each trace (red triangles) is around one order of magnitude larger than the CEP-dependent yield and scales nonlinearly with the pulse energy. It should be noted that this current does not increase monotonically, but goes through a local minimum in the field range from \SI{1.25}{\volt\per\nano\meter} to \SI{1.6}{\volt\per\nano\meter}. This current scales differently from the CEP-dependent current, implying a different origin than the nanoantenna array. Additional investigation is required as we suspect parasitic field emission from the electrodes close to the nanoantenna array or thermal emission processes play a role. We believe that an improved electrode design would greatly suppress the charge offset. Similar behavior of the charge offset has been observed in a different size antenna network shown here in the supplementary in Sec. 4.2.

\subsection{Background Charge Signal}
\begin{figure}[h!]
    \centering
    \includegraphics[width=10cm]{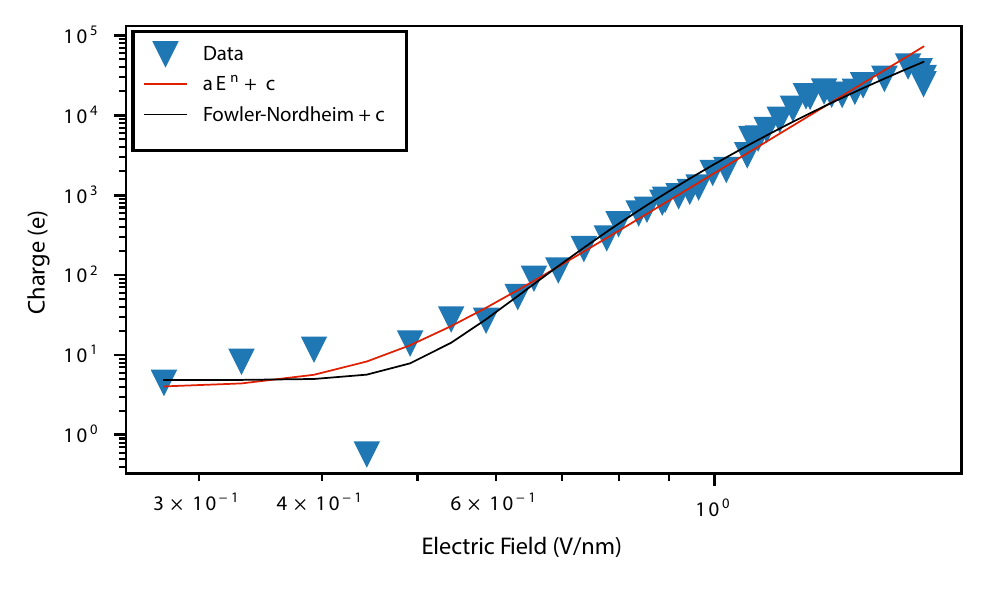}
    \caption{\textbf{Background Charge Yield:} The average charge yield as a function of peak electric field. A multi-photon absorption and Fowler-Nordheim based model are fitted to the data. For the multi-photon the function $Q(E,n,\alpha)=\alpha\cdot E^n + c$ is used, with fit results $\alpha=1.9\cdot 10^3$, $n=7.4$ and $c=3.9$. For the Fowler-Nordheim fit the function $\Gamma(E,g,\alpha)=\alpha (gE)^2 \exp{\left(-\frac{78.7}{\vert gE\vert}\right)}+c$, with the results $\alpha = 1705$, $g=15.4$ and $c=4.9$.
    \label{figS:Background}}
\end{figure}

During the measurements across all tested devices, we observed an intensity-dependent charge background. To investigate the possible origin of this contribution, we tested two different hypotheses. First, that the charge signal generated by a multiphoton emission process, and second that it is generated by a field emission process based on Fowler-Nordheim tunneling. The results are shown in Fig. \ref{figS:Background}. The multiphoton fit is defined as,
\begin{equation}
Q(E,n,\alpha)=\alpha\cdot (E)^n +c,
\end{equation}
with the polynomial order $n$ and a scaling prefactor $\alpha$. $n = 7.4$, $\alpha = 1.9\cdot 10^3$ and the offset $c=3.9$. This hypothesis implies that a 3-4 photon process is causing electron emission, which is incompatible with the photon energy of the optical pulse spanning 0.3~eV to 0.6~eV and the workfunction of gold with 5.1~eV. This means that either another process is inducing a current other than electron emission from gold or that multiphoton is not the right explanation.
The second tested hypothesis is that of a field emitter other than the nanoantenna array. To test this we used the Fowler-Nordheim fit function $\Gamma(E)$,
\begin{equation}
    \Gamma(E,g,\alpha)=\alpha (gE)^2 \exp{\left(-\frac{78.7}{\vert gE\vert}\right)} + c,
\end{equation}
with the prefactor $\alpha$, the field enhancement $g$ and critical field strength of \SI{78.7}{\volt\per\nano\meter}. The Fit results show a prefactor of $\alpha = 1705$, a field enhancement of $15.4$ and the offset $c=4.9$. This result implies that there is different field emitter causing this charge yield, as the designed field enhancement of the nanoantenna is on the order of $\sim 8$.
However, as this is merely a quantitative speculation, further research is warranted to uncover the cause of this charge contribution. One test experiment could be to use an identical device but excluding the nanoantenna arrays. With that, all contributions from the large gold leads, if also contributing, can  be measured independently. Second the change of polarization could be tested, as these nanoantennas are highly polarization sensitive \cite{putnam_optical-field-controlled_2017}. However, continuous changing of polarization in the MIR is difficult due to a lack of suitable achromatic waveplates and cannot easily be implemented.

\subsection{Larger Area Network}
\begin{figure}[ht]
    \centering\includegraphics{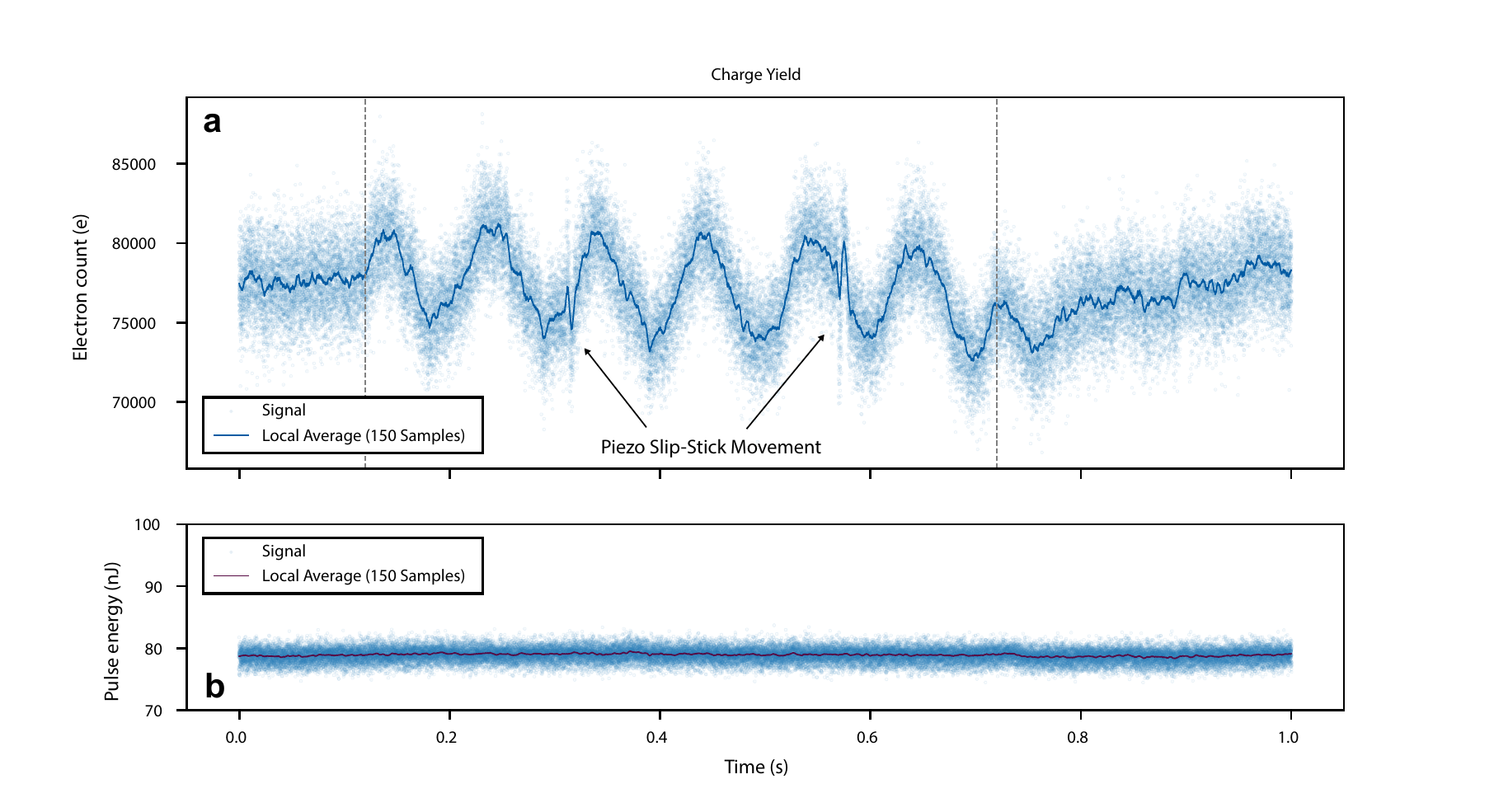}
    \caption{\textbf{Single-shot charge readout:} A single dataset recording of 50~000 laser shots for the charge yield of the nanoantenna detector (\textbf{a}) and the laser energy recorded by the pyroelectric detector (\textbf{b}). The peak field of the incident laser pulse on the array is \SI{1.6}{\volt\nano\meter}. From \SI{120}{\milli\second} to \SI{720}{\milli\second} the CE phase is linearly ramped over 6 cycles.}
    \label{fig:suppsingleshot}
\end{figure}

To verify the results measured in the main text we repeated the same measurements with an antenna network that measures \SI{30}{\micro\meter} by \SI{30}{\micro\meter}, which is substantially larger than the FWHM beam width of $\sim$~\SI{21}{\micro\meter}. The other difference between these measurements is the use of a different detector for the single-shot pulse energy, which is, in this case, a commercial mercury cadmium telluride detector.
This amounts to roughly 1000 antennas within the spatial FWHM contributing to the measured charge yield. A single-shot measurement is shown in Fig. \ref{fig:suppsingleshot}. Identical to the main text, a clear CEP modulation is present in the measured data. Furthermore, also the same signatures of the piezo slip-stick motion are present in the data. In addition, we see a 3x larger background charge signal compared to the other measurement.

As in the main text, the same frequency analysis of the single-shot measurement is shown, which presents reproducible behavior. Aside from the clear CEP peak, we see in addition the same $1/f^{3/4}$ noise characteristic is present in the data. 
The narrow band noise peaks at $\sim$ \SI{17}{\kilo\hertz} in the electron amplitude are clearly discernible from noise. In conjunction with the higher background charge signal, we can strengthen the argument that the noise peaks are driven by high frequency laser intensity changes predominantly modulating the background charge signal. 
\begin{figure}[ht]
    \centering
    \includegraphics[width=9cm]{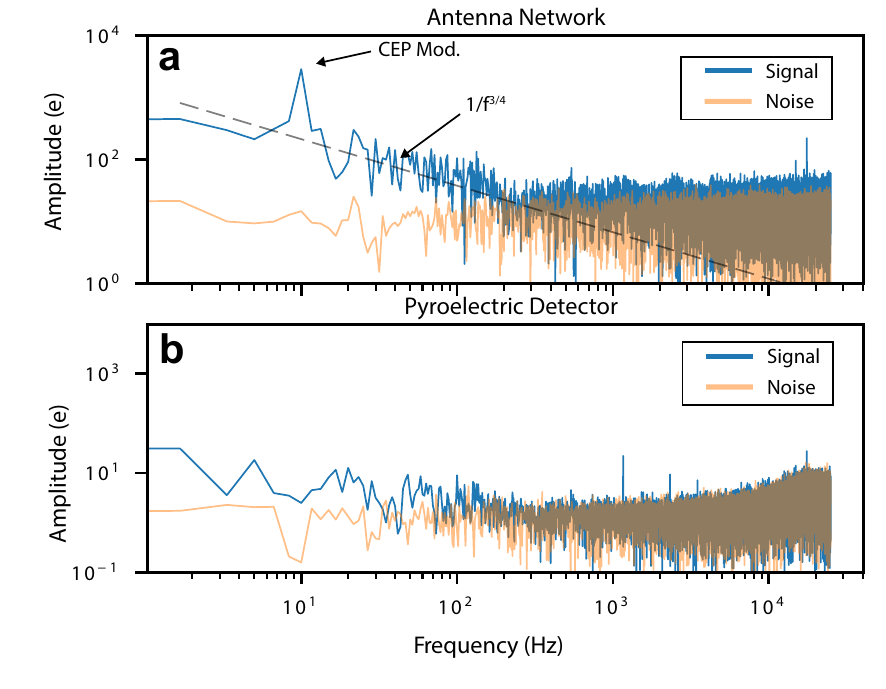}
    \caption{\textbf{Frequency Domain of the single-shot data:} The respective data from Fig. \ref{fig:suppsingleshot} $t=~$\SI{370}{\milli\second} to $t=$~\SI{620}{\milli\second} is Fourier transformed and shown in charge amplitude as a function of frequency. For comparison, the electronic noise floor is shown in orange for both spectra. (\textbf{a}) the frequency-resolved signal of the nanoantenna network. (\textbf{b}) the frequency-resolved energy signal, as a function of pyroelectric charge yield.}
\label{fig:Suppfrequencydomain}
\end{figure}

\begin{figure}[ht]
    \centering
    \includegraphics{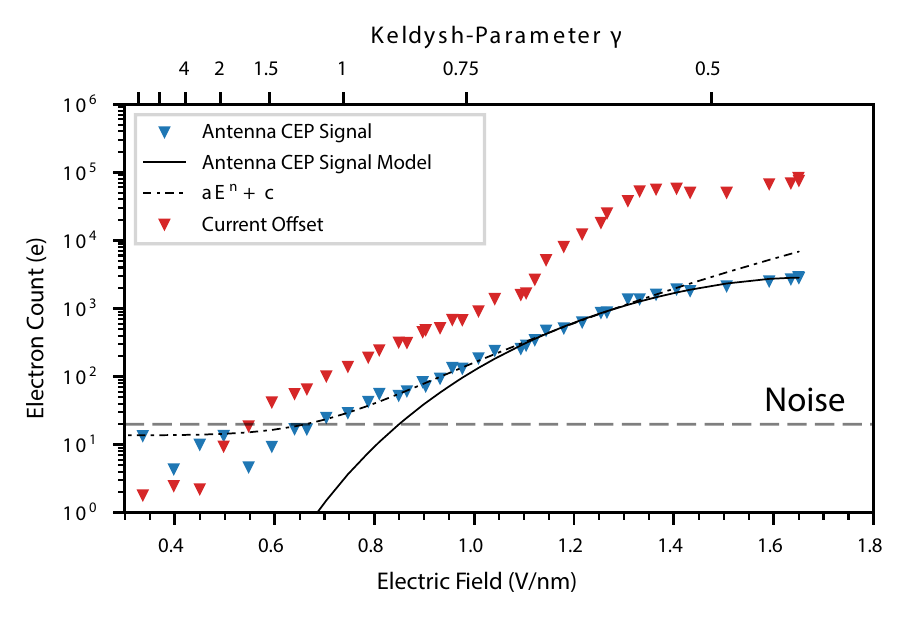}
    \caption{\textbf{Charge yield scaling:} The CE phase modulation amplitude and the average charge yield are plotted as a function of the average peak field for the respective dataset. The CE phase signal is taken from the amplitude of the \SI{10}{\hertz} frequency component of the measurement data. The antenna CE phase signal model uses the model described by Eq. 3. Furthermore, a power law fit, $aE^n+c$ to the first 30 values is shown. }
    \label{fig:Suppscaling}
    
\end{figure}
Analysing the scaling of the CEP peak as a function of incident peak field, we find in general similar behavior in Fig. \ref{fig:Suppscaling}. The Fowler-Nordheim fit results in a field-enhancement $g= 8$, very close to the simulated field-enhancement of 8.2. Furthermore the prefactor $a = 1647$ is almost identical to the one in the main text with $a'= 1517$, indicating that this array, despite the larger size, has a comparable amount of antennas contributing to the charge signal, as $a$ is proportional to the amount of antennas. Further measurements with different array sizes could map more precisely on how many antennas are involved in the charge signal. The heuristic power law fit, $Q(E) = aE^n+c$, shows as well comparable behavior to the array discussed in the main text. The fit results are $a = 147$, $n = 7.65$ and $c=13$, which in particular with the power law order $n$ agrees very well with the main text where $n'= 7.85$ is measured. This shows that the measurements are in general of predictable behavior. Although the scaling law for low field-strengths is not fully explained, models like the Yudin-Ivanov \cite{yudin_nonadiabatic_2001}, could help to explain these scaling behaviors.


\clearpage
\printbibliography